\documentclass[fleqn,useAMS,usenatbib]{mnras}
 \usepackage{amsmath}
  \usepackage{graphicx}
  \usepackage{subfigure}
  \usepackage{comment}
  \usepackage{bm}
  \usepackage{amssymb}
  \usepackage{scalerel}
  \usepackage{esvect}
  \usepackage{xcolor}
   \usepackage{upgreek}
   \usepackage[flushleft]{threeparttable}
   \usepackage{booktabs}
   \usepackage{soul}
\newcommand{\kepler}{{\em Kepler}}
\newcommand{\corot}{{\em CoRoT}}
\newcommand{\numax}{\mbox{$\nu_{\rm max}$}}
\newcommand{\Dnu}{\mbox{$\Delta \nu$}}

\newcommand{\muHz}{\mbox{$\mu$Hz}}

\newcommand{\Dpi}{\Delta \Pi}

\newcommand{\msun}{\!{\rm M_{\sun}}}
\newcommand{\np}{{n_{\rm p}}}

\title[Impact of Buoyancy Spike]
{Evolution of Dipolar Mixed-mode Coupling Factor in Red Giant Stars: Impact of Buoyancy Spike}
\author[C.~Jiang, et al.]
{\parbox{\textwidth}{C.~Jiang,$^{1}$\thanks{E-mail:jiangc@mps.mpg.de}
M.~Cunha$^{2}$, J.~Christensen-Dalsgaard$^{3}$, Q.~S.~Zhang$^{4,5,6,7}$, L.~Gizon$^{1,8,9}$}\vspace{0.4cm}\\
\parbox{\textwidth}{
$^{1}$Max-Planck-Institut f{\"u}r Sonnensystemforschung, Justus-von-Liebig-Weg 3, 37077 G{\"o}ttingen, Germany \\
$^{2}$Instituto de Astrof\'{\i}sica e Ci\^{e}ncias do Espa\c{c}o, Universidade do Porto, CAUP, Rua das Estrelas, 4150-762 Porto, Portugal\\
$^{3}$Stellar Astrophysics Centre, Department of Physics and Astronomy, Aarhus University, Ny Munkegade 120, DK-8000 Aarhus C, Denmark\\
$^{4}${Yunnan Observatories, Chinese Academy of Sciences, 396 Yangfangwang, Guandu District, Kunming 650216, China}\\
$^{5}${Center for Astronomical Mega-Science, Chinese Academy of Sciences, 20A Datun Road, Chaoyang District, Beijing 100012, China}\\
$^{6}$Key Laboratory for the Structure and Evolution of Celestial Objects, Chinese Academy of Sciences, 396 Yangfangwang, Guandu District, Kunming 650216, China\\
$^{7}${University of Chinese Academy of Sciences, Beijing 100049, China}\\
$^{8}${Institut f\"ur Astrophysik, Georg-August-Universit\"at G\"ottingen, Friedrich-Hund-Platz 1, 37077 G\"ottingen, Germany }
\\
$^{9}${Center for Space Science, NYUAD Institute, New York University Abu Dhabi, Abu Dhabi, UAE}}}

\pagerange{\pageref{firstpage}--\pageref{lastpage}} \pubyear{2019}

\def\LaTeX{L\kern-.36em\raise.3ex\hbox{a}\kern-.15em
    T\kern-.1667em\lower.7ex\hbox{E}\kern-.125emX}

\begin{document}

\label{firstpage}

\maketitle

\begin{abstract}
Mixed modes observed in red giants allow for investigation of the stellar interior structures. One important feature in these structures is the buoyancy spike caused by the discontinuity of the chemical gradient left behind during the ﬁrst dredge-up. The buoyancy spike emerges at the base of the convective zone in low-luminosity red giants and later becomes a glitch when the g-mode cavity expands to encompass the spike. Here, we study the impact of the buoyancy spike on the dipolar mixed modes using stellar models with different properties. We ﬁnd that the applicability of the asymptotic formalisms for the coupling factor, $q$, varies depending on the location of the evanescent zone, relative to the position of the spike. Signiﬁcant deviations between the value of $q$ inferred from ﬁtting the oscillation frequencies and either of the formalisms proposed in the literature are found in models with a large frequency separation in the interval 5 to 15 $\muHz$, with evanescent zones located in a transition region which may be thin or thick. However, it is still possible to reconcile $q$ with the predictions from the asymptotic formalisms, by choosing which formalism to use according to the value of $q$. For stars approaching the luminosity bump, the buoyancy spike becomes a glitch and strongly aﬀects the mode frequencies. Fitting the frequencies without accounting for the glitch leads to unphysical variations in the inferred $q$, but we show that this is corrected when properly accounting for the glitch in the ﬁtting.
\end{abstract}


\section{Introduction}

The wealth of long, near-uninterrupted, high-precision asteroseismic data acquired by space missions, such as \corot\ \citep{baglin06} and \kepler\ \citep{borucki10}, allow us to investigate the inner structure of stars by analysing the observed dipolar mixed modes (modes with angular degree $\ell = 1$) that have a mixed pressure-gravity nature. Mixed modes are extensively detected in subgiant and red-giant stars, in which the local gravitational acceleration, and hence the buoyancy frequency, increases due to the extreme condensation of the core. This opens up the possibility of the mixing between pressure and gravity waves. Mixed modes in stars on the subgiant branch (SGB) were first observed and modelled in $\eta$ Boo \citep{ jcd95, hans95a, hans03, gun96, car05} and $\beta$ Hyi \citep{bed07, bra11}, while those on the red-giant branch (RGB) have been widely analysed with \corot\ and \kepler\ data \citep[e.g.,][]{hek09,bed10,hub10,jiang11,mat11,mos11,bau12,kal12}. The gravity-mode (g-mode) character of mixed modes carries physical information about the core region. By determining the g-mode period spacing ($\Dpi$) of mixed modes in red giants it was possible to probe the structure of the core \citep{bed11}, and identification of frequency splittings allowed monitoring its rotation \citep{bec12,mos12a}, for the first time. The studies of the period spacing benefited from large-scale measurements \citep{datta15, vrard16} of long-duration observations. On the other hand, the pressure-mode (p-mode) character allows the waves to penetrate the outer part of stars and hence become detectable \citep{jcd14}. 

The mixture of the p and g modes can be measured by the dimensionless coupling factor ($q$) that is related to the size of the evanescent zone (EVZ) between the p-mode cavity in which oscillations are present with pressure as restoring force, and the g-mode cavity in which buoyancy is the restoring force. In the EVZ, oscillations have an exponential behaviour and thus are evanescent \citep{bookas}. The coupling factor has proven to be a useful seismic diagnostic of the inner structure of RGB stars by many studies of mixed modes \cite[e.g.,][]{mos12a, jiang14, hek17, mos17, pincon20}. With the assumption that $q$ varies little with mode frequency for detectable modes, measurements of $q$ were first done for \kepler\ evolved RGB stars \citep{mos12b,buy16} by fitting the asymptotic expression of the frequency pattern of mixed modes, derived by \citet[][hereafter, S79]{shiba79}, to the observations. Based on the same asymptotic expression, \citet[][hereafter, J20]{jiang2020a} measured $q$ for individual mixed modes of RGB models in different evolutionary stages and found that the frequency-dependence of $q$ is not negligible for RGB stars, which was also reported in previous works \citep{jiang14,hekker18,cunha19}.   

The asymptotic formalism of $q$ from S79 was derived under the assumption that the EVZ is very thick and hence that the coupling is weak, as is expected in evolved RGB stars. However, this is not the case for low-luminosity giant stars near the transition between SGB and RGB as well as in the red clump \citep{mos17} in which the coupling is strong. An important breakthrough was made by \citet[][hereafter, T16]{takata16} who paid special care to the strong-coupling case. T16 derived an asymptotic formalism for $q$ for dipolar mixed modes using the JWKB method (Jeffreys, Wentzel, Kramers, and Brillouin), and taking the perturbation to the gravitational potential into account. In that way, the derived asymptotic $q$ can take large values close to 1 (but not beyond) in the limiting case of very strong coupling. \cite{hekker18} compared $q$ that was calculated using the asymptotic formalism from T16 with that obtained by fitting the mode frequencies following the formalism proposed by \cite{jcd12} and further developed by \cite{jiang14}, \cite{cunha15} and \cite{hek17}, which is an alternative description of the asymptotic expression used by \cite{mos12b}. \cite{hekker18} found that the values of the asymptotic $q$ are significantly smaller compared to the fitted values for evolved RGB models where the frequency at maximum power ($\numax$) is low, which was also reported by an earlier work of \cite{mos17}. J20 monitored the evolution of $q$ obtained by fitting the asymptotic expression of mixed modes to theoretical frequencies, and found that $q$ generally decreases with stellar evolution for early- and intermediate-age RGB models, which is followed by a dramatic drop for more evolved RGB models. This is related to the structure of the mid-layers that are located between the hydrogen-burning shell and the neighbourhood of the base of the convective zone (BCZ) of red giants. The relation between $q$ and the mid-layers of red giants in different stages was extensively investigated by \cite{pincon20}. They showed that the values of $q$ depend on the size of the EVZ and the local density scale height. Specifically, the broadening of the EVZ is mainly responsible for the observed decrease in $q$ in evolved red giants with mean large frequency spacings $\Dnu \lesssim 10\, \muHz$. Moreover, \cite{pincon20} further pointed out that the asymptotic formalism derived either from S79 or T16 is not the best solution for a medium coupling case where the EVZ is between very thin and thick. 

An important feature of the mid-layer structure is the structural variation 
resulting from the evolution of the convective envelope.
In subgiant and early red-giant stars the convective envelope deepens in mass with age in the so-called {\it first dredge-up}, reaching regions whose composition has been modified by nuclear reactions.
In models without diffusion this gives rise to a discontinuity in the hydrogen abundance and hence density, causing a spike in the density gradient and thus the buoyancy frequency at the BCZ.
With further evolution, the convective envelope retracts in mass, leaving behind
the composition discontinuity and the spike in the radiative region and hence potentially in the g-mode cavity.
As the hydrogen-burning shell approaches the discontinuity and begins to feel the lower mean molecular weight above the discontinuity, the luminosity temporarily decreases with evolution, leading to {\it the red-giant bump}, until the shell reaches and eliminates the discontinuity \citep{jcd15}. 
If the spike is in the g-mode cavity it has the character of a so-called \textit{buoyancy glitch} that can significantly affect the mixed-mode properties during the bump phase \citep{cunha15}.
The mode frequencies and inertias change, and consequently, so does the value of the inferred $q$ if the impact of the glitch is not explicitly accounted for.
However, in low-luminosity RGB stars, much before the star reaches the luminosity bump, this spiky structural variation (sharp and small scale) emerges in the buoyancy frequency just at the BCZ and hence outside the g-mode cavity for observable oscillation modes. Although the buoyancy spike is not in the oscillatory cavity and thus has little impact on the properties of observed mixed modes, it indeed affects the characteristic frequencies in the mid-layers that are crucial for the asymptotic analysis of mixed modes. Neither the glitch nor the small-scale spiky structure was considered in the analysis of \cite{pincon20}. Therefore, a comprehensive study of $q$ is needed for those RGB stars that are earlier than RGB bump, to account for the presence as well as the impact of the buoyancy spike.

In this paper, we monitor the evolution of the coupling factor with theoretical models which exhibit a spike in the buoyancy frequencies, aiming at investigating the relation between the coupling factor and stellar interior structure. We focus on the regions near the base of the convective zone, exploring an interpretation that accounts for the discrepancy between the asymptotic and the fitted coupling factor. The paper is organised as follows. Section~\ref{sc:asy} presents the framework of our analysis. We use the asymptotic formalisms from S79 and T16 to calculate asymptotic values for $q$ which are then compared with the fitted $q$ obtained from J20. A discussion concerning the evolution of $q$ is provided in Section~\ref{sc:qevo}, emphasising the impact of the buoyancy spike. In Section~\ref{sc:glitch} we discuss specifically the impact of the buoyancy glitch on the fitted $q$ for models approaching close to the bump, fitting the glitch in the way following \cite{cunha15, cunha19}. Finally, a summary is given in Section~\ref{sc:conclusion}.

\section{Asymptotic Analysis of Mixed Modes}
\label{sc:asy}

\subsection{Asymptotic Formalisms}
\label{sc:qasy}

The oscillations detected in evolved red-giant stars are associated with a very thick EVZ, as normally assumed by previous works \citep{tassoul80,unno89}, and thus with the case of very weak coupling. According to S79 the coupling factor in the weak-coupling case is asymptotically given by
\begin{equation}
q_0 = \frac{1}{4} \exp \left(-2 \uppi X_I \right),
\label{eq:weakq}
\end{equation}
where $X_I$ is an integral term that is made over the layers encompassing the EVZ (defined in Appendix~\ref{app:x}).
The value of $q_0$ is thus in theory between 0 and 0.25, and is strongly related to the width of the EVZ, which is determined by the mode frequency and the two characteristic frequencies: the \textit{Brunt-V\"ais\"al\"a frequency} ($N_{\rm BV}$, also called the \textit{buoyancy frequency}) and the \textit{Lamb frequency} ($S_\ell$, $\ell$ being mode degree) \footnote{Note that the modified versions of the two characteristic frequencies (equation~\eqref{eq:modified}) are used here in the definition of $X_I$. Nevertheless, we still refer to equation~\eqref{eq:weakq} as S79 throughout the paper.}. A thicker EVZ generally produces weaker coupling for the mixed modes, hence a smaller value of $q$. However, in the opposite limiting case of a very thin EVZ, equation~\eqref{eq:weakq} is no longer valid to estimate the coupling factor.

\begin{figure}
\resizebox{\columnwidth}{!}{\includegraphics[angle =0]{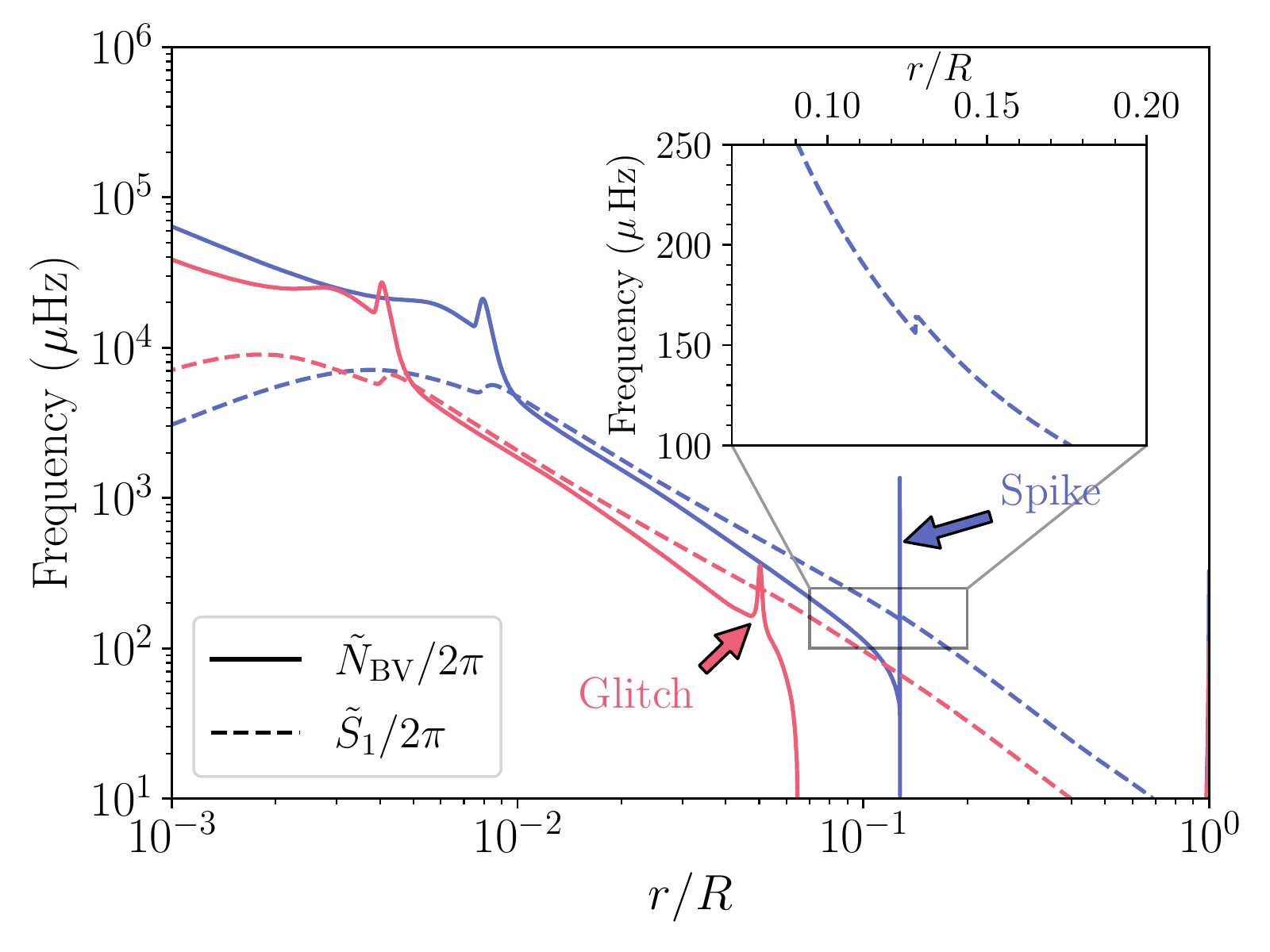}}
\caption{Propagation diagram showing $\tilde{N}_{\rm BV}$ and $\tilde{S}_{\rm 1}$ for two 1.0~$\msun$ models with calibrated solar metallicity ($Z=0.0173$), one early RGB (Model 1, blue, $\numax = 204.51 ~\muHz$, $\Dnu = 16.27~\muHz$), and one more evolved RGB (Model 2, red, $\numax = 60.41~\muHz$, $\Dnu = 6.35~\muHz$) model. The $\tilde{N}_{\rm BV}$ are indicated by solid curves and the $\tilde{S}_1$ by dashed curves. The sharp variations (marked by the arrows) in the buoyancy frequency of both models are caused by the chemical discontinuity. Locations of the two models in the HR diagram are illustrated in Figure~\ref{fg:track}. The insect shows specifically for Model 1 the tiny jump in $\tilde{S}_{\rm 1}$ that is also due to the chemical discontinuity.}
\label{fg:propagation}
\end{figure}

T16 considered the limiting case of a very thin EVZ, hence very strong coupling, which is usually the case observed in subgiants and young red giants as well as in clump stars. T16 fully accounted for the effect of the perturbation to the gravitational potential that is neglected by the Cowling approximation \citep{cowling41} in S79, and introduced an asymptotic formalism for the coupling factor in such strong-coupling cases. The formalism by T16 uses a modified version of the two characteristic frequencies introduced by \cite{tak06}, to avoid the Cowling approximation, which are defined as
\begin{equation}
 \tilde N_{\rm BV} = \frac{N_{\rm BV}} {J} \,\,\, \text{and} \,\,\,  \tilde S_1 = J S_1,
 \label{eq:modified}
\end{equation}
where $S_1$ is the normal Lamb frequency for $\ell=1$ modes and $J$ is defined as
\begin{equation}
J = 1 - \frac{1}{3} \frac{\mathrm{d} \ln M_r}{\mathrm{d} \ln r} = 1 - \frac{4}{3} \frac{ \uppi \rho r^3} {M_r},
\label{eq:j}
\end{equation}
with $M_r$ being the concentric mass contained within radius $r$. The average density of the concentric mass $M_r$ is larger than the density $\rho$ at radius $r$, except at the centre. As a result, $M_r$ is always larger than $\frac{4}{3}\uppi \rho r^3$ and hence $J$ is positive but never exceeds unity\footnote{When $J=1$, equation~\eqref{eq:modified} reproduces the normal version of the two characteristic frequencies, thus assumes the Cowling approximation.}. Therefore,  $\tilde S_1$ and $\tilde N_{\rm BV}$ would produce a slightly narrower EVZ than their normal counterparts do. In Figure~\ref{fg:propagation} $\tilde N_{\rm BV}$ and $\tilde S_1$ are shown for two 1.0 $\msun$ RGB models: one young model (Model 1) much below the RGB bump, and one evolved model (Model 2) closer to the bump (locations of the models in the HR diagram are illustrated in Figure~\ref{fg:track}). In principle, acoustic waves travel in an outer cavity where the mode angular frequency $\omega > (\tilde S_1$ and $\tilde N_{\rm BV}$), while gravity waves propagates in a central radiative cavity below the convective envelope where $\omega < (\tilde N_{\rm BV}$ and $\tilde S_1$). The EVZ, where the mode behaviour is exponential, is located between the profiles of $\tilde N_{\rm BV}$ and $\tilde S_1$. 
The integral region of $X_I$ is determined by the interfaces between the propagative regions and the EVZ, where we have $\omega = \tilde S_1$ and $\omega = \tilde N_{\rm BV}$ that are referred to as the turning points with fractional radius being $x_1$ and $x_2$, respectively. According to T16's new formalism, the asymptotic value of the coupling factor in the case of a thin EVZ is defined by
\begin{equation}
q_1 = \frac{1 - \sqrt{1-\exp(-2 \uppi X)}} { 1 + \sqrt{1-\exp(-2 \uppi X)}}.
\label{eq:qasy}
\end{equation}
The variable $X$ is connected to stellar interior structure and is defined in T16 as
\begin{equation}
X = X_{I} + X_{R},
\label{eq:x}
\end{equation}
in which the term $X_R$ is closely connected with the gradient of the two characteristic frequencies near the centre of the EVZ (see Appendix~\ref{app:x} for details). The turning points $x_1$ and $x_2$ are frequency dependent. The centre of the EVZ, $x_{\rm c}$, is defined in the sense of T16 as $x_{\rm c} = \sqrt{x_1 x_2}$, and hence is also frequency dependent.

Compared to equation~\eqref{eq:weakq}, the new term $X_R$ in equation~\eqref{eq:x} is important in the case of a thin EVZ, otherwise $X$ is dominated by the integral term $X_I$ for a thick EVZ. $X$ is positive and never gets to exactly 0 because of the positive $X_R$, and therefore $q_1$ is always less than unity. The asymptotic limit of equations~\eqref{eq:weakq} and~\eqref{eq:qasy} converges as $X_{I} \to \infty$, corresponding to extremely weak coupling seen in evolved RGB stars. However, a comparison between the formalisms provided by equations~\eqref{eq:weakq} and~\eqref{eq:qasy}, as well as a study regarding their applicability to real RGB stars or stellar models with different types of EVZ are still nonexistent, and are, thus, required. In addition, these asymptotic formalisms assume that the background structure of the star is smooth over scales comparable to the local wavelength (e-folding length) of the mode in the propagation region (evanescent zone). Therefore, they are expected to fail in cases where the sharp structural variations are present in the buoyancy frequency, as those shown in Figure~\ref{fg:propagation}. It is then of interest to understand if either $q_0$ or $q_1$ computed from the models are still a good representation of the coupling inferred from fitting the frequencies in these cases. If not, a formalism that corrects the asymptotic analysis for the presence of the spike is thus needed.

\subsection{Method}
\label{sc:method}

The assessment of the buoyancy spike's impact on mixed modes is done by comparing the asymptotic $q_{\rm asy}$ with the fitted $q_{\rm fit}$ for RGB models with different properties and ages. Here $q_{\rm asy}$ is derived from the formalisms of equations~\eqref{eq:weakq} and~\eqref{eq:qasy} presented in Section~\ref{sc:qasy}, while $q_{\rm fit}$ is obtained from J20 by fitting the theoretical dipole-mode cyclic frequencies ($\nu = \omega/2\uppi$) with the formalism introduced by \cite{mos12b},
\begin{equation}
 \nu = \nu_{\np} + \frac{\Dnu}{\uppi}\arctan \left[q_{\rm fit} \tan \uppi \left(\frac{1}{\nu \Dpi_{\rm asy}}-\epsilon \right) \right],
 \label{eq:mixnu}
 \end{equation}
where $\Dpi_{\rm asy}$ is the asymptotic value of the g-mode period spacing, $\nu_{\np}$ is the uncoupled solution for the p mode with radial order $\np$ and the gravity offset $\epsilon$ is linked to the properties of the EVZ in RGB stars. 

In this analysis, we continue using the models discussed in J20. In total around 500 models were computed with different initial masses ($M=1.0-1.8~\msun$ with a 0.2~$\msun$ step) and three chemical abundances ($Z = 0.0099$, 0.0173 and 0.0295), covering a fraction of the RGB on each evolutionary track, starting from the transition between SGB and RGB to a point close to the luminosity bump (see Figure~\ref{fg:track} for an illustration of the $1~\msun$ evolutionary track). The models were generated with the Aarhus STellar Evolution Code \citep[{\scriptsize ASTEC},][]{jcd08a} and the corresponding oscillation frequencies with the Aarhus adiabatic oscillation package \citep[{\scriptsize ADIPLS},][]{jcd08b}. Convection was treated under the assumption of the mixing-length theory \citep{boh58} and a mixing-length parameter of 1.96 obtained from a calibration to solar models was fixed for all the models. For other input physics we refer to Section~3 of J20, but we note that overshooting, diffusion, and rotation were not considered in the models.
Both equations~\eqref{eq:weakq} and~\eqref{eq:qasy} are used to derive $q_{\rm asy}$ for all RGB models selected along the evolution tracks. The fitting procedure for $q_{\rm fit}$ was introduced in Section~4 of J20 with the Bayesian-nested-sampling tool \textsc{Diamonds} \citep{corsaro14}.


\section{Evolution of the Coupling Factor}
\label{sc:qevo}

From an analytical study of the mixed modes and the EVZ, \cite{pincon20} concluded that the value of the coupling factor depends in general on the thickness of the EVZ as well as on the local density scale height. In this section, we first discuss the frequency dependence of the coupling factor with the two RGB models illustrated in Figure~\ref{fg:track}, using the asymptotic formalisms given in Section~\ref{sc:qasy}, and explain how different regions of the EVZ are related to the mixed modes properties. Then we verify the applicability of the asymptotic formalisms by comparing the asymptotic and the fitted values of the coupling factor. Finally, we conclude on the evolution of the coupling factor for low-luminosity RGB stars. 

\subsection{Frequency Dependence of $q_{\rm asy}$}
\label{sc:fre_qasy}

The frequency dependence of the coupling factor has been confirmed and highlighted in previous works \citep[e.g.,][]{jiang14, mos17, hekker18, cunha19,jiang2020a, pincon20}.  Here we compute the asymptotic values of $q_{\rm asy}$ using the formalisms given in Section~\ref{sc:qasy} to illustrate how the frequency dependence of $q_{\rm asy}$ is related to the stellar structure.
To do that, we regard the relevant quantities as continuous functions of frequency,
with the understanding that only the oscillation mode frequencies have
physical meaning.
To probe fully the behaviour we consider frequencies in a frequency range with a width of more than $10 \, \Dnu$
($\numax$
\footnote{For models, $\numax$ is estimated according to the scaling relation given by equation (10) in \cite{kje95}, in terms of surface gravity and effective temperature.}
is included but not necessarily centred in this range). The younger model (Model 1, $\Dnu = 16.27\, \muHz$) in Figure~\ref{fg:track} is adopted for the structure configuration shown in Figure~\ref{fg:q_vs_fre}. As illustrated in panel A of the figure, $\tilde N_{\rm BV}$ and $\tilde S_1$ are plotted to exhibit the variations of the EVZ, as well as the buoyancy spike structure that coincides with the BCZ at this relatively early stage of RGB. The asymptotic values for $q_{\rm asy}$ are obtained using equations~\eqref{eq:weakq} and~\eqref{eq:qasy}, for $q_0$ and $q_1$, respectively.
In panel B of Figure~\ref{fg:q_vs_fre}, $q_0$ and $q_1$ for Model 1 are plotted against frequency, showing two brutal jumps in $q_1$ at around 110 and 160~$\muHz$, respectively, which are closely related to the buoyancy spike. 
The property of the EVZ is inevitably affected by this structural variation. 
The older model (Model 2, $\Dnu = 6.35~\muHz$) is shown in Figure~\ref{fg:q_vs_fre1} for the case of buoyancy glitch. The analysis for each model uses the same grid as in the frequency calculations of J20, with more than 25000 meshpoints distributed between the centre and the surface. This is set up to guarantee adequate spatial resolutions of the models, in particular in the region of the sharp structural feature. In models such as Model 1, where the convective envelope encompasses an increasing mass fraction of the star, the composition and density discontinuities are replaced consistently by a steep and fully resolved change over a very small range in radius. In models with a retracting convective envelope, such as Model 2, the discontinuity is smoothed somewhat by numerical diffusion in the evolution calculation.

\subsubsection{Evanescent Zone}
\label{sc:evz}

\cite{pincon20} considered models in different evolutionary stages and classified the corresponding EVZ according to three categories: thick and thin EVZ, and a transition between the two. The EVZ in SGB and young RGB stars extends in a thin layer somewhere between the hydrogen-burning shell and the BCZ. For evolved RGB stars, $\numax$ and the detectable oscillation modes become low enough for the EVZ to be located in the convective zone and its lower bound to be very close to the position of the BCZ, denoted in fractional radius as $x_{\rm bcz}$. At this point the EVZ is expected to be much thicker. A thick EVZ corresponds to the weak-coupling regime under which equation~\eqref{eq:weakq} was derived, while a thin EVZ corresponds to the strong-coupling regime under which equations~\eqref{eq:qasy} and~\eqref{eq:x} were derived. For the transition region, it is not clear whether weak or strong coupling should be considered. Nevertheless, \cite{pincon20} only considered the models in which the buoyancy spike is absent. Moreover, no spike was considered in either of the asymptotic formalisms, which assume a smooth background.

The RGB models analysed here are mostly young ones, with $x_{\rm bcz} \approx x_{\rm spike}$, with $x_{\rm spike}$ being the position of the buoyancy spike. As illustrated in Figure~\ref{fg:q_vs_fre} for the case of Model 1, the EVZ is located in one of three different regions if the buoyancy spike is taken into consideration. For high-frequency modes, the rather thin EVZ is located in the radiative zone and so are the turning points ($x_2 < x_1 < x_{\rm spike}$). We define such an EVZ as being located in \textit{Region\,A}. As the frequency decreases, the EVZ moves out toward the convective zone. The first jump in $q_1$ emerges at $\sim 160\,\muHz$ when the outer turning point $x_1$ passes the spike but the centre of the EVZ ($x_{\rm c}$) is still in the radiative zone $( x_{\rm spike} \leq x_1  ~ \text{and} ~x_{\rm c} \leq x_{\rm bcz})$. We define the EVZ of such modes as the transition EVZ and being located in \textit{Region\,T}. The second jump in $q_1$ follows at $\sim 110\,\muHz$ as $x_{\rm c}$ also passes the BCZ ($x_{\rm bcz} < x_{\rm c} < x_1$); this occurs because $X_R$ depends specifically on structural quantities at $x_{\rm c}$ (see Appendix~\ref{app:x}). At this point the outer half of the EVZ is within the convective zone and we define such an EVZ as being located in \textit{Region\,B}. For even lower-frequency modes, with $x_2$ also past $x_{\rm bcz}$, the entire EVZ is inside the convective zone and they are also in Region\,B.
The classification of the EVZ is summarised in Table~\ref{tb:evz}. Figure~\ref{fg:q_vs_fre1} shows the case for Model 2 for which the buoyancy spike has moved inside the radiative core, so that $x_{\rm spike} < x_{\rm bcz}$; thus, it is located inside the g-mode propagation cavity for oscillation frequencies around $\numax$. The presence of the buoyancy spike and glitch induces more complex variations in $q_{\rm asy}$ in this case, but we apply the same dividing criteria of the EVZ for evolved models like Model 2, for the reason explained in Section~\ref{sc:qfre}.

We note that our division of EVZ regions is independent of the thickness of the EVZ. For instance, EVZ in Region\,A can be thin for early RGB stars but relatively thick for evolved ones. However, EVZ in Region\,B is unambiguously thick. In Region\,T it is rather hard to determine whether the EVZ is thick or thin.

\subsubsection{Impact of the Buoyancy Spike on $q_{\rm asy}$}
\label{sc:qfre}

Panel B of Figure~\ref{fg:q_vs_fre} highlights the comparison between $q_0$ and $q_1$ for the EVZ located in different regions. The value of $q_0$ is mainly determined by the size of the EVZ which grows as the mode frequency decreases. Therefore, $q_0$ generally decreases from Region\,A to Region\,B, but shows a subtle jump at the inner edge of Region\,T when the outer turning point $x_1$ is just past the spike and the EVZ starts to encompass the little sharp variation in the $\tilde S_1$ that is also caused by the chemical discontinuity (see an illustration for Model 1 in the inset of Figure~\ref{fg:propagation}). As for $q_1$, its sensitivity to the gradient term $X_R$ dominates the variations in each region. In Region\,A the characteristic frequencies $\tilde S_1$ and $\tilde N_{\rm BV}$ show similar radial variations that can be approximated by a power law with similar exponent \citep{mos17, pincon20}, so $X_R$ barely changes with frequency in the region (see panel C of Figure~\ref{fg:q_vs_fre}). This is true until the outer turning point reaches the buoyancy spike. At this point, the above mentioned tiny jump in $\tilde S_1$ together with the much more significant sharp variation in $\tilde N_{\rm BV}$ contribute to the first drop in $X_R$ at $\sim160\,\muHz$. At the outer edge of Region\,T, the buoyancy spike coincides with $x_{\rm c}$, where $X_R$ is derived, which induces dramatic variations in the inferred $X_R$ as well as $q_1$. The fact that $\tilde S_1$ and $\tilde N_{\rm BV}$ vary differently with radius when $x_{\rm c}$ is past the BCZ also contributes with a smaller scale variation to $X_R$ at the outer edge of Region\,T. The EVZ in Region\,B enlarges drastically, hence the coupling factor drops quickly there. In Region\,B, $q_1$ is gradually dominated by the integral term $X_I$, and it eventually converges with $q_0$ at the extremely weak-coupling case.

When the star evolves approaching the luminosity bump, the spike structure also moves inward, which induces extra jumps to $q_1$ inside Region\,T. Figure~\ref{fg:q_vs_fre1} illustrates such configurations of the EVZ regions for Model 2. However, $\numax$ in the evolved RGB models like Model 2 is so low that the detectable modes only have a Region\,B EVZ. Hence, all the models with a Region\,T EVZ analysed here indeed resemble what is shown in Figure~\ref{fg:q_vs_fre}, meaning that the buoyancy spike only affects the $q_{\rm asy}$ in young RGB models.
Furthermore, the employed definition of the three regions in our analysis ensures that the significant perturbations and jumps of $q_{\rm asy}$ are only seen in Region\,T, thus leaving the other two regions unaffected by the sharp structural variation.

In summary, as mode frequency decreases, $q_0$ decreases rather smoothly except for a tiny jump, but $q_1$ experiences sharp variations near the edges of Region\,T. It is interesting to see that in Region\,A and Region\,B, $q_0$ is greater than $q_1$, which is due to the additional gradient term $X_R$ that lowers the value of $q_1$.
The two converge as the thickness of the EVZ becomes extremely large. Equations~\eqref{eq:weakq} and~\eqref{eq:qasy} are derived based on different limiting cases in terms of EVZ sizes. But the ambiguous determination of the EVZ's thickness and the presence of the structural spike, not accounted for in the asymptotic analysis, brings into question the applicability of these asymptotic formalisms in each of the three regions.

\subsection{Applicability of the Asymptotic Formalisms}
\label{sc:qasy_qfit}

Figures~\ref{fg:q_vs_fre} and~\ref{fg:q_vs_fre1} illustrate the EVZ for two RGB models. We have conducted similar analyses for other models in different evolutionary stages, such that the models possess different kinds of EVZ according to the three regions introduced in Section~\ref{sc:evz}. In general, detectable oscillation modes in the SGB and young RGB stars are found to have an EVZ in Region\,A, while evolved RGB stars have one in Region\,B, and an EVZ in the transition Region\,T exists at some point between young and evolved RGB. 

With the three possible regions for the EVZ defined, we now discuss which one of the asymptotic formalisms (equations~\eqref{eq:weakq} and~\eqref{eq:qasy}) is more applicable in each region by comparing $q_{\rm asy}$, containing $q_0$ and $q_1$, with $q_{\rm fit}$ from J20. 
The method used in J20 to obtain $q_{\rm fit}$ is also applicable to observed mixed modes \citep[e.g.][]{hekker18,jiang18}, and thus $q_{\rm fit}$ should be close to what is expected for real stars with similar properties. 
The fitting process for $q_{\rm fit}$, applying equation~\eqref{eq:mixnu}, was separately performed for each avoided crossing centred on $\nu_{\np}$, using modes with frequencies in the range [$\nu_{\rm p-m} - \Dnu / 2, \nu_{\rm p-m} + \Dnu / 2$] for each run, with $\nu_{\rm p-m}$ being the frequency of the most p-mode-like mode that is essentially very close to $\nu_{\np}$. By utilising equations~\eqref{eq:weakq} and~\eqref{eq:qasy} $q_{\rm asy}$ can be computed as a continuous function of frequency, but for the purpose of comparison with $q_{\rm fit}$, here $q_{\rm asy}$ is also derived at each $\nu_{\np}$.
Bearing this in mind, both asymptotic formalisms are assessed in each region with a $\chi^2$ test. The $\chi^2$ function is defined as
\begin{equation}
\chi^2 = \frac{1}{N} \sum_{i=1}^N \left( \frac{q_{{\rm asy},i} - q_{{\rm fit},i} }{\sigma_i} \right)^2,
\end{equation}
where $N$ is the total number of modes analysed, and the uncertainty $\sigma$ is set to be 0.02, the typical uncertainty of the coupling factor measured in early RGB stars \citep{mos17,hekker18}. 

\begin{figure*}
\resizebox{1.0\hsize}{!}{\includegraphics{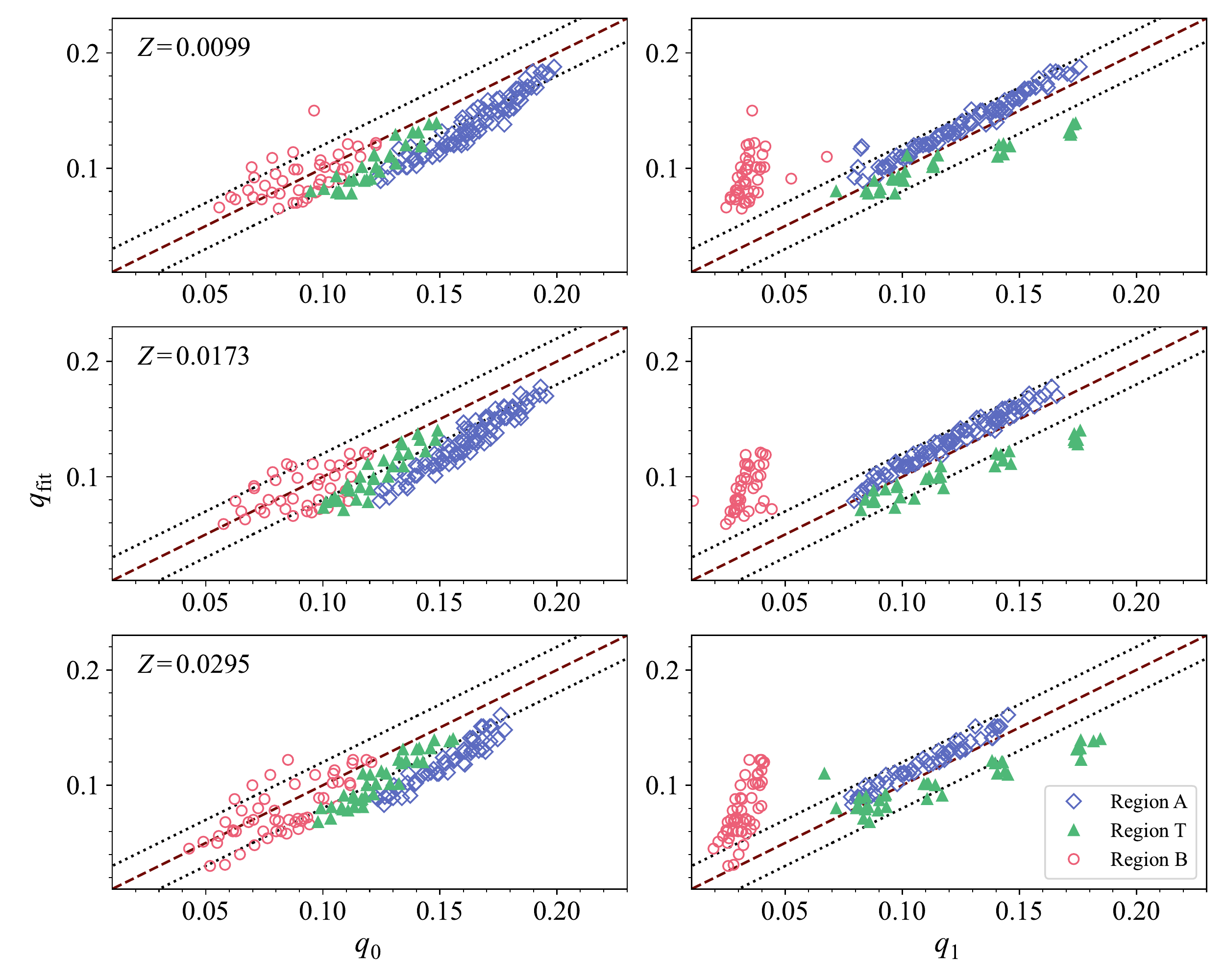}}
\caption{Coupling factor ($q_{\rm fit}$) computed by fitting the model frequencies as in J20 vs. the asymptotic coupling factor computed using equations~\eqref{eq:weakq} and~\eqref{eq:qasy} for the same models as $q_{\rm fit}$. Only observable modes with $\np$ in the range 10 to 13 (whose frequencies are near $\numax$) are presented. Symbols indicate different types of evanescent zone: Region\,A (blue open diamond), Region\,T (green filled diamond) and Region\,B (red open circle). Plots in the same row present models sharing the same initial metallicity. \textit{Left column}: the asymptotic coupling factors are computed using equation~\eqref{eq:weakq} for all the models and denoted as $q_0$ . \textit{Right column}: the asymptotic coupling factors are computed using equation~\eqref{eq:qasy} for all the models and denoted as $q_1$.
The dashed line indicates agreement and the dotted lines show the adopted 1\,$\sigma$ error of 0.02.} 
\label{fg:qdif}
\end{figure*}

To show the differences of the two formalisms, we apply them to all the models with EVZ in the different regions. In the following we generally focus on modes expected to be observationally detectable, in the vicinity of the estimated $\numax$.
Specifically, we often consider frequencies in the range $\numax \pm 3 \Dnu$. However, in following the evolution with age of mode properties it is more convenient to fix the acoustic-mode order $\np$; for this alternative we choose modes with $10 \le \np \le 13$, which roughly correspond to the above range of frequency.
Comparisons for all the models are illustrated in Figure~\ref{fg:qdif}. In Region\,A, regardless of metallicity, $q_1$ shows a better agreement with $q_{\rm fit}$ than $q_0$, as expected for a relatively thin EVZ. The agreement is within the adopted 1\,$\sigma$ error.
On the other hand, in Region\,B $q_0$ matches $q_{\rm fit}$ much better than $q_1$.
Things become much more complicated in Region\,T, due to the presence of the buoyancy spike. Generally speaking, $q_0$ reproduces $q_{\rm fit}$ well in the Region\,T in cases with higher $q$ values while $q_1$ shows alignment with $q_{\rm fit}$ for lower $q$ cases. This is contrary to the expectation of the asymptotic theory that $q_0$ works better in a thick EVZ and $q_1$ is more suitable for a thin EVZ. The results evidently call for a correction to the asymptotic formalisms for the presence of the buoyancy spike in this transition region, which is beyond the scope of this analysis. 

The comparison is quantified in Table~\ref{tb:chi2}, where we list the $\chi^2$ values computed for each region, which supports what is depicted in Figure~\ref{fg:qdif}. It can be seen that when $q_1$ is used to compare with $q_{\rm fit}$ the values of $\chi^2$ are always smaller in Region\,A, while they are much larger in Region\,B. Moreover, from a crude visual check, $q_1 \sim 0.12$ seems to be the threshold between the ``higher and lower'' $q$ scenarios for Region\,T. The value of $\chi^2$ is substantially lowered when it is computed with $q_0$ for models with $q_1 > 0.12$ and with $q_1$ for the rest of the models. The reduced $\chi^2$ values are denoted as Region\,T$_{\rm q}$ in Table~\ref{tb:chi2}.   

\begin{figure*}
\resizebox{1.0\hsize}{!}{\includegraphics{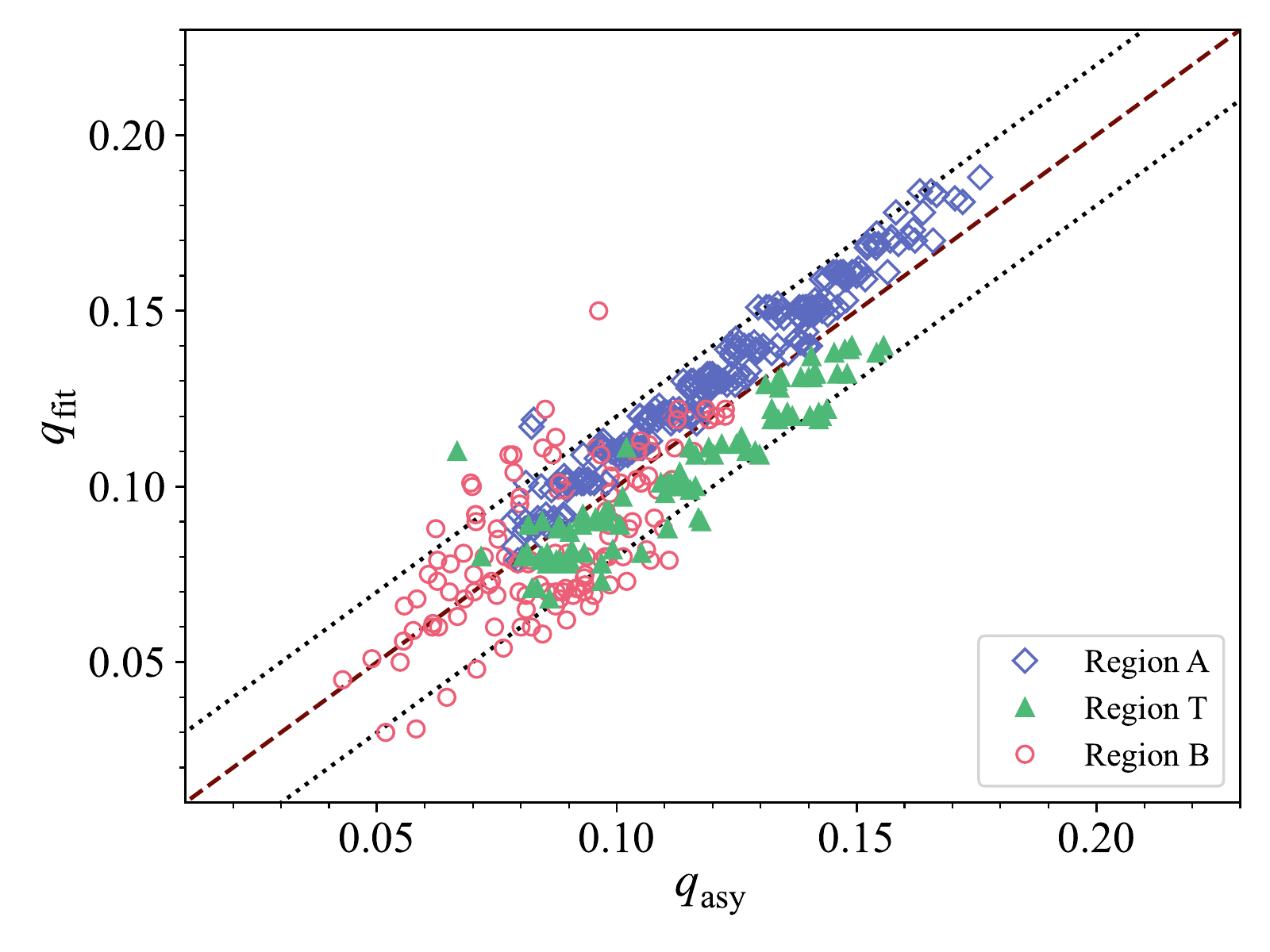}}
\caption{As in Figure~\ref{fg:qdif}, comparisons for observable modes, defined as to have $10 \le \np \le 13$. 
The values of $q_{\rm asy}$ are derived as: $q_1$ for Region\,A and Region\,T models with $q_1 < 0.12$, $q_0$ for Region\,B and Region\,T models with $q_1 > 0.12$ (see main text).} 
\label{fg:qall}
\end{figure*}

Therefore, from the comparisons presented in Figure~\ref{fg:qdif} and Table~\ref{tb:chi2} we conclude that equation~\eqref{eq:qasy} from T16 represents the fitted coupling factor better in Region\,A, while equation~\eqref{eq:weakq} from S79 is more applicable in Region\,B. 
In both these cases the buoyancy spike plays no significant role and, thus, the asymptotic expectations are  fulfilled within the limits in which they were derived. 
For models in the transition region, it is harder to decide which asymptotic formalism is more suitable. From a crude examination of the fitting agreement from Figure~\ref{fg:qdif}, we see that for Region\,T models, equation~\eqref{eq:weakq} and equation~\eqref{eq:qasy} can be used to fit for mixed modes with $q_1 < 0.12$ and those with $q_1 > 0.12$, respectively. 
According to the results presented in the next section,  Region\,T models are found with $\Dnu$ in the interval 5 to 15\,$\muHz$.
In Figure~\ref{fg:qall} a comparison between $q_{\rm fit}$ and $q_{\rm asy}$ computed according to these rules for the asymptotic formalisms is shown for all modes with $\np$ between 10 and 13.

\subsection{Evolution of $q_{\rm asy}$}
\label{sc:evo_asy}

The evolution of $q_{\rm fit}$ was examined by J20 for thousands of mixed modes obtained in RGB models. The authors found that $q_{\rm fit}$ decreases as the model evolves on the RGB and the decreasing trend is dependent on the radial order ($\np$) of the p-mode component of mixed modes: the decrease in $q_{\rm fit}$ with time can be very fast for low-$\np$ modes, while it is generally gradual for high-$\np$ modes in the detectable frequency range. 
In this section we discuss the evolution of the coupling factor in terms of the variation of $q_{\rm asy}$ for the same mixed modes analysed in J20, to shed some light on how and when the spike begins to affect the mixed modes. The values of $q_{\rm asy}$ are computed according to the applicability rules of the asymptotic formalisms established in Section~\ref{sc:qasy_qfit}.   

The evolution of $q_{\rm asy}$ as a function of $\nu_{\np}$ at different ages is shown for different stellar masses in Figure~\ref{fg:q_evo_173} for modes with $\np$ between 10 and 13. The model ages increase from right to left in each diagram so that the red symbols are the oldest ones. The modes with the same $\np$ are connected by the same line and are equally spaced in age. For models with masses $\leq 1.6 ~ \msun$, the age difference between two connected modes is 0.02$\,$Gyr. For the $1.8~\msun$ models, an age interval of  0.01 Gyr is used, to ensure an adequate number of models along the evolution. 

\begin{figure*}
\resizebox{1.0\hsize}{!}{\includegraphics[angle =0]{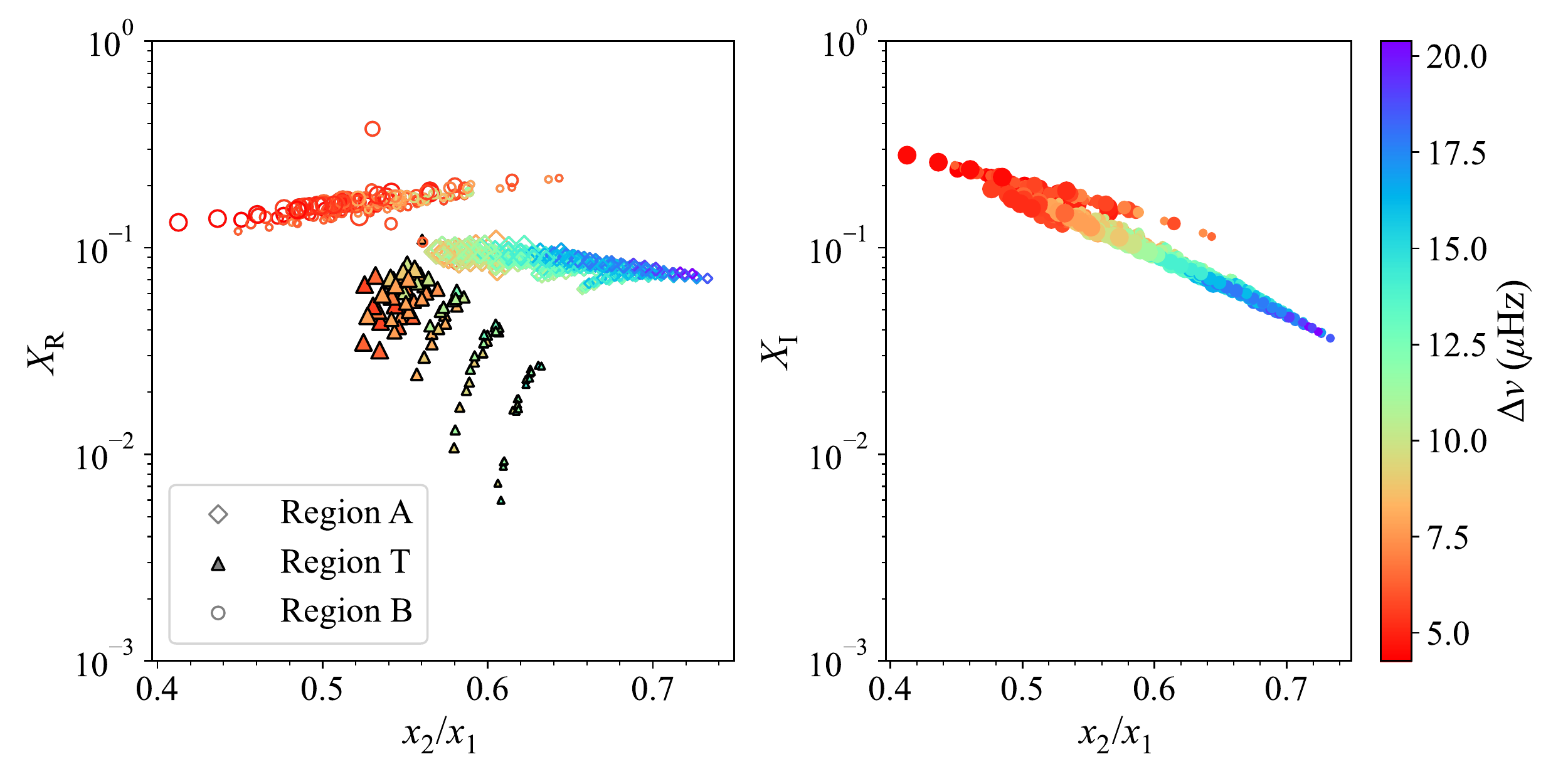}}
\caption{$X_R$ for modes in the interval $10 \le \np \le 13$ as a function of $x_2 / x_1$, calculated for all our models. Symbols are colour coded according to $\Dnu$, and the symbol size is proportional to the model mass.
For comparison purposes, values of the integral term $X_I$ are shown in the right plot.
The ratio $x_2 / x_1$ decreases as the model evolves from right to left.} 
\label{fg:xr}
\end{figure*}

Similar to J20, a general descending trend of the asymptotic coupling factor is found for these high-$\np$ modes. However, in every case a small peak is seen for intermediate-age models (green- and yellowish ones) that have an EVZ in Region\,T. As discussed in Section~\ref{sc:qfre}, this small increase in $q_{\rm asy}$ is attributed to the insufficient treatment of the buoyancy spike in both asymptotic formalisms used to calculate $q_{\rm asy}$ when the EVZ of the model enters and leaves Region\,T. That maximum is followed by a decrease in $q_{\rm asy}$ which is due to the significant increase in the width of the EVZ throughout Region\,B. 
Examining the variation of $X_R$ as a function of the location ratio $x_2/x_1$ for all the modes, as illustrated in Figure~\ref{fg:xr}, gives an idea on how the regions of the EVZ are related to the mid-layers of RGB stars.
In general younger models possess a thin EVZ and hence a large $x_2/x_1$ value close to 1. The ratio decreases as the EVZ broadens with stellar age.
The $X_R$ values of these Region\,A young models grow as $x_2/x_1$ decreases when $x_2/x_1 \gtrsim 0.55$, while in Region\,B evolved models, in contrast, $X_R$ shows a decrease with decreasing $x_2/x_1$ when $x_2/x_1 \lesssim 0.65$. The overlapping regime of $ 0.5 \lesssim x_2/x_1 \lesssim 0.65$ is Region\,T, showing a sharp drop in $X_R$, corresponding to a $\Dnu$ spread between 5 to 15\,$\muHz$. Moreover, $X_I$ behaves as expected, in that the thinner the EVZ the smaller the $X_I$, but one may still observe a subtle displacement in the value of this quantity in the Region\,T regime. A detailed discussion of the contributing factors to the variations of $q_{\rm asy}$, either from the sharp structural variation in Region\,A or the dramatic enlargement of EVZ in Region\,B, is needed, which should be done in future investigations.

In addition, it is expected that $x_2/x_1 < 1$ in these low-luminosity RGB models, but at SGB or very early RGB we note that the ratio may be larger than unity. In such situation, the model is so young that the EVZ would be in Region\,A. The youngest 1.0 $\msun$ models indeed have an $x_2/x_1 > 1$ for very low-frequency modes and according to \cite{takata16} equation~\eqref{eq:qasy} still holds for the swapped loci of $x_1$ and $x_2$.

\section{Impact of Buoyancy Glitch}
\label{sc:glitch}	  

As the young red-giant stars evolve toward the bump, their hydrogen-burning shell located outside the core approaches the chemical discontinuity that is left behind by the retreating convective envelope during the first dredge-up.
From previous discussions, we now know that the chemical discontinuity leads to sharp buoyancy variations that emerge as the spike structure at the BCZ in early RGB stars (see Figure~\ref{fg:propagation}).
According to Section~\ref{sc:qevo} the spike has a strong impact on the $q_{\rm asy}$ for relatively young RGB models that have an EVZ in Region\,T. Observationally, the coupling parameters are affected by the spike only in RGB stars that are much closer to the luminosity bump. In these more evolved stars the convective envelope retreats to include the sharp structural variation in the g-mode cavity. Since the width of the buoyancy spike is much shorter than the local wavelength, the spike is sufficiently sharp to cause significant deviations of the mixed-mode properties from their simple asymptotic behaviour \citep{cunha15, cunha19} and becomes a buoyancy glitch.
The observed frequency pattern and the properties inferred from the fit with equation~\eqref{eq:mixnu}, meaning the inferred period spacing and the coupling factor, are also inevitably affected, as confirmed by J20 when they tried to obtain $q_{\rm fit}$ and $\Dpi$ of mixed modes while deliberately neglecting the impact of the glitch in several RGB bump models. 

\begin{figure}
\resizebox{\columnwidth}{!}{\includegraphics[angle =0]{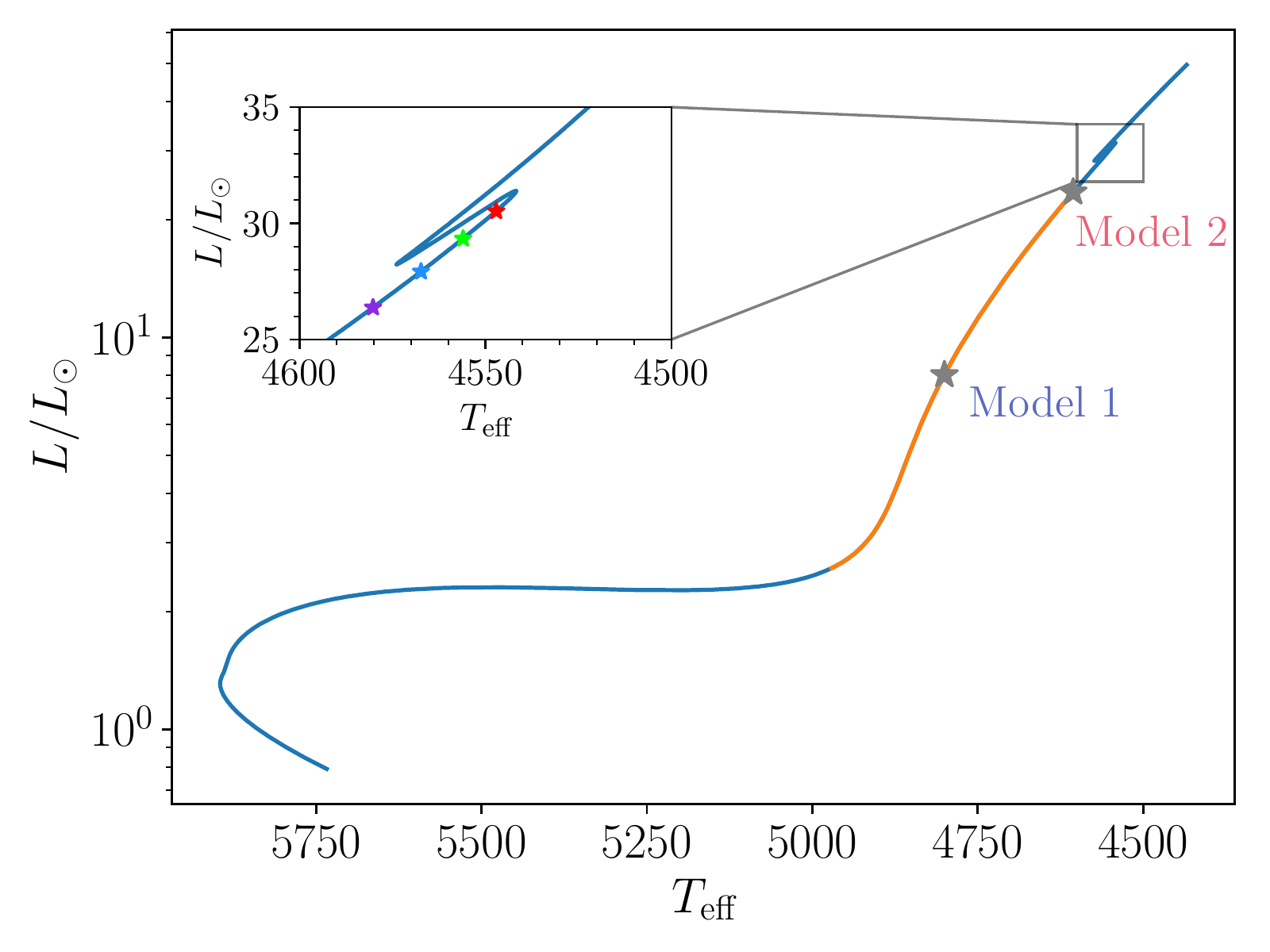}}
\caption{
Evolution track of the 1.0 $\msun$ and solar composition model. Two models with their $\tilde{N}_{\rm BV}$ and $\tilde{S}_{\rm 1}$ presented in Figure~\ref{fg:propagation} are highlighted by the star symbol. The section of models considered in Section~\ref{sc:qevo} are covered in orange. A similar coverage of the track is also employed for other models with different masses and metallicities. The inset shows the section of the evolution track around the luminosity bump and four evolved models analysed in Section~\ref{sc:glitch}.}
\label{fg:track}
\end{figure}

\begin{figure*}
\resizebox{1.0\hsize}{!}{\includegraphics[angle=0]{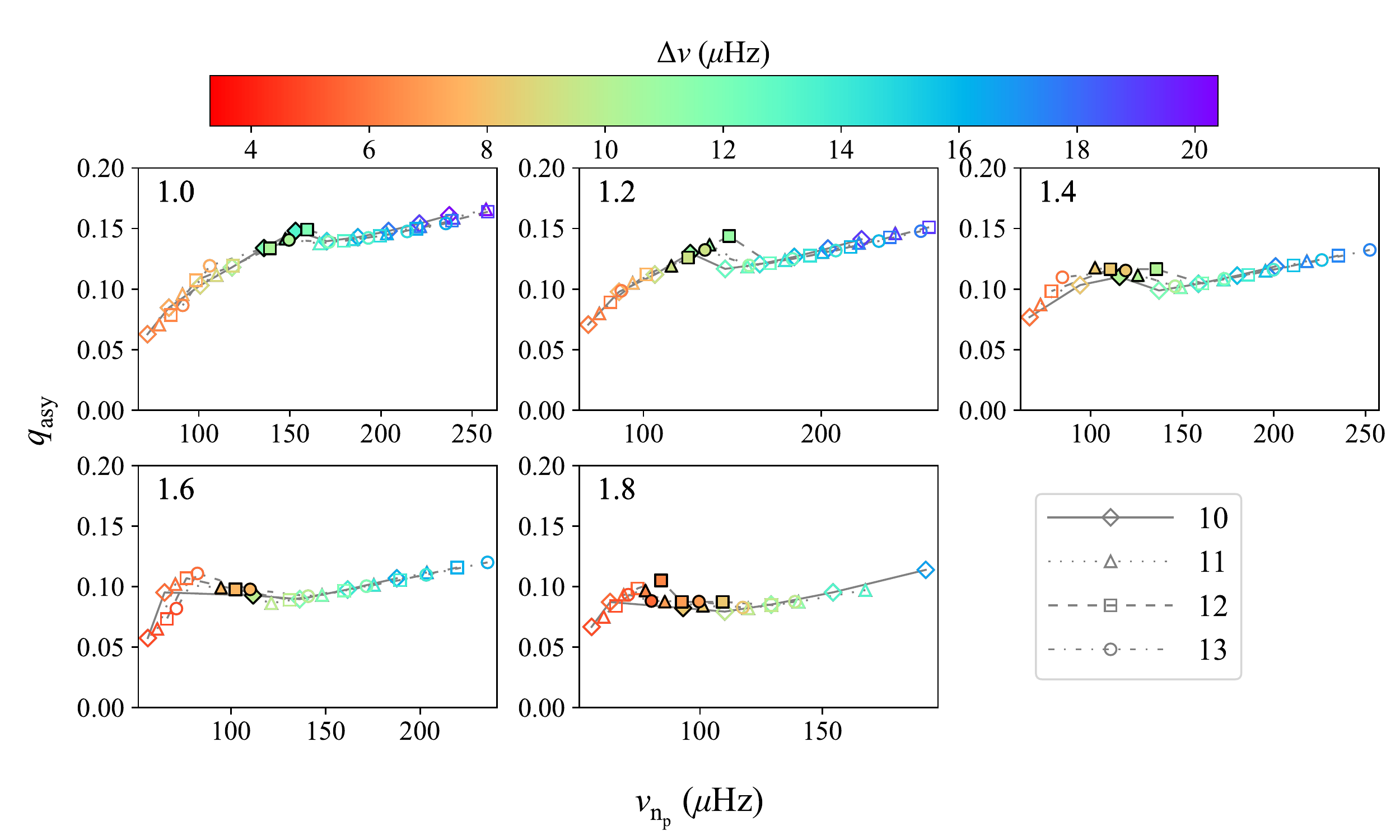}}
\caption{Evolution of $q_{\rm asy}$ for modes in the interval $10 \le \np \le 13$ in models with different masses (values shown in units of solar mass in the upper left corner of each diagram) and $Z = 0.0173$. The acoustic resonance frequency $\nu_{\np}$ of an $\np$-order mode decreases with model age, so the evolution goes from higher to lower $\nu_{\np}$. Model ages are also indicated by the colours coded according to $\Dnu$. The models are plotted equally spaced in age (details of the age spacing given in Section~\ref{sc:evo_asy}). Modes of the same $\np$ are indicated by the same symbol and linked by the same line. Notations of symbols and line types for each $\np$ are given in the legend. Region\,T modes are indicated by filled symbols.} 
\label{fg:q_evo_173}
\end{figure*}

According to \cite{cunha15}, for low-mass stars the glitches can be observed at the RGB luminosity bump, including a small region before the bump as well as the region where the luminosity decreases. However, J20 noticed that the mixed modes are significantly influenced by the glitch only for models below the bump when the luminosity is still rising. 
This could be partially due to the fact that the glitch becomes broader than the local wavelength in the J20 models after the bump, as a result of the numerical diffusion intrinsic to {\scriptsize ASTEC}. Moreover, very soon after the bump the buoyancy spike disappears as it is reached by the nuclear-burning shell that is moving out in mass as the helium core grows. At the hydrogen-burning shell the chemical-composition variation also leads to a localised maximum in the buoyancy frequency, but this feature is much broader than the local wavelength of the oscillation and thus has no effect on the mixed modes. 
Therefore, here we discuss the glitch-induced impact on the coupling factor with models located before the bump. Specifically, four 1~$\msun$, $Z=0.0173$ RGB models closely approaching the bump are selected for the analysis (Fig.~\ref{fg:track}). Their buoyancy frequencies with glitches are presented in the left column of Figure~\ref{fg:glitch}, from which we can see that the glitch migrates inward in radius from the BCZ as the model evolves. The impact of the glitch on $q_{\rm fit}$ of dipolar mixed modes\footnote{These four models are older than those shown in Figure~\ref{fg:q_evo_173}. Thus, here the highest-frequency mode has an $\np = 12$.} with $\np$ between 1 and 12 is shown in the top panel of Figure~\ref{fg:qglitch}, for the four models. We can see that $q_{\rm fit}$ is perturbed by the glitch, with the perturbation starting to show in lower-frequency (lower-$\np$) modes for the youngest model (the purple one) and becoming significant also in higher-frequency (higher-$\np$) modes as the star evolves.

\citet[][hereafter C19]{cunha19} presented an analytical representation of the dipole-mode period spacing in the presence of a glitch, which is expressed as 
\begin{equation}
\frac{\Dpi}{\Dpi_{\rm asy}} = \left[ 1 - \mathcal{F}_{\rm G,C} \right]^{-1},
\label{eq:glitch}
\end{equation}
where the term $\mathcal{F}_{\rm G,C}$ reflects the impact of the glitch as well as the mode coupling. Here the values of the asymptotic period spacing $\Dpi_{\rm asy}$ are obtained from J20 and are equivalent to their $\Dpi_1$. Detailed definitions of $\mathcal{F}_{\rm G,C}$ and other related quantities are given by equations~(24) to (26) in C19. 
Following the two-step approach employed by C19, 
here we obtain the coupling factor $q^\prime_{\rm fit}$ included in $\mathcal{F}_{\rm G,C}$ by fitting equation~\eqref{eq:glitch} to the period spacings computed from {\scriptsize ADIPLS}. We adopt again \textsc{Diamonds} to do Bayesian sampling and estimate $q^\prime_{\rm fit}$ through the fitting to {\scriptsize ADIPLS} period spacings.
Firstly, the glitch properties (Table~\ref{tb:glitch_para}) of the models are derived by fitting the artificially computed frequencies of what would be the pure g modes if no coupling existed for the four RGB models. 
In this step, the fit is performed with a similar expression to equation~\eqref{eq:glitch} but discarding the impact of coupling (as explained in Section 3 of C19).
The glitch-impacted period spacings obtained in this step are illustrated by the red curves in the right column of Figure~\ref{fg:glitch}.
Secondly, the fit is redone, but including the additional impact of the coupling to determine $q^\prime_{\rm fit}$, using equation~\eqref{eq:glitch} with the glitch parameters fixed to the values inferred in the previous step.
Following C19, in the second step we consider that the frequency-dependence of coupling factor is represented through a linear relation,
\begin{equation}
q^\prime_{\rm fit} = q^\prime [\alpha (\nu / \numax - 1) + 1],
\label{eq:linq}
\end{equation}
where the pair of parameters ($q'$, $\alpha$) characterises the mode coupling. Equation~\eqref{eq:linq} can be easily transferred to the normal linear model as
\begin{equation}
q^\prime_{\rm fit} = \beta \nu + c
\label{eq:linq1}
\end{equation}
with
\begin{align}
\beta &= q^\prime \alpha / \numax, \\
c &= - q^\prime \alpha + q^\prime.
\end{align}
Again, in order to be compared with $q_{\rm fit}$ and $q_{\rm asy}$, $q^\prime_{\rm fit}$ is also estimated at each $\nu_\np$.
With the glitch parameters obtained from the first step fixed and $\nu_\np$ replacing $\nu$ in equation~\eqref{eq:linq}, the second step of the fitting process is performed to determine $q'$ and $\alpha$.
The estimated parameters of the best-matching models are listed in Table~\ref{tb:glitch}. $\Dpi$ derived from equation~\eqref{eq:glitch} using these parameters and the glitch-related ones, are compared with the {\scriptsize ADIPLS} period spacings in Figure~\ref{fg:glitch}. The parameters reproduce well the deviations of $\Dpi$ induced by the glitch and the coupling, though there are a few discrepancies at the minima of the spacings, which can be reduced by taking the pure p-mode frequencies as free parameters in the fitting procedure (see C19). However, this would greatly increase the number of free parameters and hence the difficulty of convergence, but would not improve significantly the inferred $q^\prime_{\rm fit}$ which is the parameter of interest in this work. Furthermore, from the slope coefficient $\beta$ of the linear model listed in Table~\ref{tb:glitch}, we can see that the dependence of the coupling factor on the mode frequencies increases for evolved models, which emphasises again that the frequency dependence cannot be neglected especially when analysing the oscillation spectra of evolved RGB stars \citep{pincon20}. Finally, a comparison between $q^\prime_{\rm fit}$ derived from the best-fiting models and glitch-impact-free $q_{\rm asy}$ ($q_0$) from equation~\eqref{eq:weakq} is illustrated in the bottom panel of Figure~\ref{fg:qglitch}. After taking the glitch-induced impact into account, $q^\prime_{\rm fit}$ matches $q_{\rm asy}$ well for detectable modes with frequencies around $\numax$, but noticeable differences exist for lowest and highest order modes, implying a possible better description of a quadratic polynomial model for the frequency-dependent $q$.

\section{Conclusion}
\label{sc:conclusion}

Hundreds of RGB models with different masses and metallicities have been analysed to study the coupling factor of dipolar mixed modes in red-giant stars. In young RGB stars located before the luminosity bump, the chemical discontinuity left behind by the retreating convective envelope during the first dredge-up leads to a spike signature in the buoyancy frequencies, which consequently affects the coupling parameter of the mixed modes for stars in different evolutionary stages. 

The buoyancy spike emerges in low-luminosity RGB stars as a sharp variation at the base of the convection zone and outside the g-mode cavity of detectable mixed modes. 
When the spike is present in the evanescent zone, the coupling parameter is inevitably affected compared to the spike-free evanescent zone. In this analysis, we show that the evanescent zone can be classified according to its location in three different regions namely: a spike-free, thin EVZ (located in Region\,A), a thick EVZ also free of the spike influence (located in Region\,B) and a transition EVZ, which may be thin or thick, in which the buoyancy spike has a significant impact (located in region\,T). 
With these RGB models, comparisons between the asymptotic coupling factor derived based on the  formalisms of \cite{shiba79} and \cite{takata16} and the fitted ones from equation~\eqref{eq:mixnu} assist us to better understand the applicability of the asymptotic formalisms in each region of the evanescent zone. The asymptotic formalisms generally work as expected, in that equation~\eqref{eq:qasy} from \cite{takata16} represents the fitted coupling factor better in Region\,A for the stronger coupling cases, while equation~\eqref{eq:weakq} based on \cite{shiba79} is more adequate in Region\,B, for the weaker coupling cases. However, in the transition region our $\chi^2$ analysis of the comparisons shows that there is a limit for the use of the formalisms, as expected given that both formalisms assume a smooth background. Despite this fact, from a crude examination of the fitting agreement in Figure~\ref{fg:qdif} and Table~\ref{tb:chi2} (last row), we see that equation~\eqref{eq:weakq} and equation~\eqref{eq:qasy} are still applicable in the transition region for modes with coupling factor less and greater than 0.12, respectively. 

Knowing the applicability of the asymptotic formalisms allows for further monitoring the variation of the coupling factor along the evolution of stars that have an evanescent zone in different regions. For mixed modes with frequencies in the detectable range, a general descending trend of the asymptotic coupling factor is observed, resembling the variation of the fitted coupling factor reported by \cite{jiang2020a}. Furthermore, a small increase in the asymptotic coupling factor is seen in intermediate-age models that have an EVZ in Region\,T. This is due to the disregard of the buoyancy spike in both asymptotic formalisms used to calculate $q_{\rm asy}$ in the transition layer. Our models show that for the detectable oscillations, models with an EVZ located in Region\,T have a large frequency separation in the interval 5 to 15\,$\muHz$. Then following the small increase we see quick drops in $q_{\rm asy}$ for the Region\,B models, which is also in the analysis of the fitted coupling factor \citep{jiang2020a}. 

More evolved Region\,B stars approach the RGB luminosity bump as the buoyancy spike moves inside the g-mode cavity of the observed modes and becomes a glitch that induces a significant impact on the properties of the mixed modes. If the frequencies are fitted without accounting for the glitch impact, the coupling factor inferred from observations varies irregularly with frequency. However, such variations are not expected in the asymptotic coupling factor. In fact, the asymptotic coupling factor is not affected as the glitch is outside the evanescent zone where the coupling factor is derived, for these more evolved stars. This apparent disagreement is overcome in the analysis of four RGB models that are located just before the bump. Indeed, we found that the fitted coupling factor of the mixed modes reproduces the asymptotic counterpart well, when the glitch is explicitly taken into consideration using the analytical formalism from \cite{cunha19}. Moreover, the comparison between the coupling factor inferred from the fitting with this formalism and the asymptotic value shows that there might be a quadratic dependence of the coupling factor on frequency, which becomes more evident as the model ages. 

This analysis opens the way to explore further the impact of the buoyancy spike on the properties of mixed modes in low-luminosity red-giant stars. However, we only conducted the analysis with stellar models. It will be very interesting to compare the results with observations, which shall be considered in future work. Furthermore, this work highlights that we are still lacking a formalism that corrects the asymptotic description of the coupling factor for the presence of the rapid structural variation in the evanescent zone, such as that seen in Region\,T. An analysis of that kind will be key to our understanding of the mid-layer structure of red-giant stars. 

\section*{Acknowledgements}

This work is supported by a grant from the Max Planck Society to prepare for the scientific exploitation of the PLATO mission.
M. S. Cunha is supported by national funds through Funda\c{c}\~{a}o para a Ci\^{e}ncia e a Tecnologia (FCT, Portugal) -- in the form of a work
contract (CEECIND/02619/2017) and through the research grants UIDB/04434/2020, UIDP/04434/2020 and PTDC/FIS-AST/30389/2017,
and by FEDER - Fundo Europeu de Desenvolvimento Regional through COMPETE2020 - Programa Operacional Competitividade e Internacionaliz\'{a}cao (grant: POCI-01-0145-FEDER-030389).
Funding for the Stellar Astrophysics Centre is provided by The Danish National Research Foundation (Grant DNRF106). Funding for Yunnan Observatories is co-sponsored by the Strategic Priority Research Program of the Chinese Academy of Sciences (grant No. XDB 41000000), the National Natural Science Foundation of China (grant No. 11773064), the foundation of Chinese Academy of Sciences (Light of West China Program and Youth Innovation Promotion Association), and the Yunnan Ten Thousand Talents Plan Young \& Elite Talents Project.

\section*{Data Availability}

The data underlying this article will be shared on reasonable request to the corresponding author.

\clearpage

\begin{table}
\caption{Regions of the evanescent zone.} \label{tb:evz}
\begin{tabular}{cccc}
\hline
\hline
\noalign{\smallskip}
 EVZ  &  Region\,A  & Region\,T & Region\,B \\
\noalign{\smallskip}
\hline
\noalign{\smallskip}
 Location & $x_2 < x_1 < x_{\rm spike}$ & $x_{\rm spike} \leq x_1  ~ \& ~x_{\rm c} \leq x_{\rm bcz}$ & $x_{\rm bcz} < x_{\rm c} < x_1$ \\
\noalign{\smallskip}
Formalism & T16 & S79 or T16 & S79 \\
$q_{\rm asy}$ &  $> 0.13$ & $\sim\,0.09 - 0.15$ & $\lesssim0.13$  \\
\noalign{\smallskip}
\hline
\noalign{\smallskip}
\end{tabular}
\begin{tablenotes}
\setlength\labelsep{0pt}
\scriptsize
\item \textbf{Notes.} {The classification of the EVZ is based on the locations of the turning points $x_1$ and $x_2$, the centre of the evanescent zone $x_{\rm c}$, the buoyancy spike $x_{\rm spike}$ and the base of the convective zone $x_{\rm bcz}$. S79 and T16 denote the asymptotic formalisms of the coupling factor based on \cite{shiba79} and \cite{takata16}, respectively. For Region\,T models, S79 is applicable for $q_1<0.12$ and T16 for $q_1>0.12$.}
\end{tablenotes}
\end{table}

\begin{table}
\caption{Values of $\chi^2$ showing the agreement of $q_{\rm asy}$ with $q_{\rm fit}$.} \label{tb:chi2}
\begin{tabular}{ccccccc}
\hline
\hline
\noalign{\smallskip}
 $Z$  &  \multicolumn{2}{c}{0.0099}  & \multicolumn{2}{c}{0.0173} & \multicolumn{2}{c}{0.0295} \\
\noalign{\smallskip}
\hline
\noalign{\smallskip}
	$\chi^2$   & $\chi^2(q_0)$ & $\chi^2(q_1)$ & $\chi^2(q_0)$ & $\chi^2(q_1)$ & $\chi^2(q_0)$ & $\chi^2(q_1)$ \\
\cmidrule(lr){2-3} \cmidrule(lr){4-5} \cmidrule(lr){6-7}
\noalign{\smallskip}
Region\,A & 1.85 & $0.42$ & 2.48 & $0.29$ & 2.86 & $0.25$   \\
Region\,T & 1.10 & $1.13$  & 1.27 & $1.49$ & 1.27 & $1.76$   \\
Region\,B & 0.64 & $9.04$ & 0.66 & $8.64$& 0.72 & $6.30$  \\
\hline
Region\,T$_{\rm q}$ & \multicolumn{2}{c}{0.31}  & \multicolumn{2}{c}{0.46} & \multicolumn{2}{c}{0.52}\\
\noalign{\smallskip}
\hline
\noalign{\smallskip}
\end{tabular}
\begin{tablenotes}
\setlength\labelsep{0pt}
\scriptsize
\item \textbf{Notes.} {For Region\,T$_{\rm q}$, $\chi^2$ is computed with $q_{\rm asy}$ from equation~\eqref{eq:weakq} for models with $q_1 < 0.12$ and from equation~\eqref{eq:qasy} for the rest of the models in Region\,T.}
\end{tablenotes}
\end{table}

\begin{table}
\caption{Glitch parameters derived by fitting the artificially computed frequencies of what would be the pure g modes if no coupling existed for the four RGB models. 
} \label{tb:glitch_para}
\begin{tabular}{ccccccc}
\hline
\hline
\noalign{\smallskip}
 Age  & $\Dnu$ & $\Dpi_{\rm asy}$ & $A_{\rm G}$  & $\Delta_{\rm g}$ & $\varpi_{\rm g}^{\star}$ & $\delta$\\
  (Gyr) & ($\muHz$) & (s) & ($\mu{\rm rad} \, {\mathrm s}^{-1}$) & ($\mu {\rm rad} \, {\mathrm s}^{-1}$) & ($\mu {\rm rad} \, {\mathrm s}^{-1}$) \\
\noalign{\smallskip}
\hline
\noalign{\smallskip}
11.126 & 5.65 & $69.20$  &183 & $55$  & 517  & 1.36\\
11.131 & 5.38 & $68.36$  & 202 & $74$  & 845  & 1.16\\
11.135 & 5.15 & $67.57$ & 213 & $106$  & 1436 & 1.17\\
11.139 & 4.97 & $66.81$ & 254 & $167$  & 2564  & 1.20\\
\noalign{\smallskip}
\hline
\end{tabular}
\begin{tablenotes}
\setlength\labelsep{0pt}
\scriptsize
\item \textbf{Notes.} {The glitch is characterised by three parameters, $A_{\rm G}$ and $\Delta_{\rm g}$, which are the amplitude and width of the glitch, respectively, and the glitch position $r^\star$, which determines the buoyancy depth $\varpi_{\rm g}^\star$ at the glitch position. The phase $\delta$ is related to the details of the mode reflection near the turning points of the propagation cavity. For detailed definitions of these parameters we refer to \cite{cunha15}.}
\end{tablenotes}
\end{table}

\clearpage

\begin{table}
\caption{Parameters obtained from the fit of equation~\eqref{eq:glitch} to the period spacings computed from {\scriptsize{ADIPLS}} for the four models with a core glitch. The values are from the best-matching models that are illustrated in Figure~\ref{fg:glitch}. } \label{tb:glitch}
\begin{tabular}{cccccc}
\hline
\hline
\noalign{\smallskip}
 Age  & $\Dnu$ & $q^\prime$ & $\alpha$  & $\beta$\\
  (Gyr) & ($\muHz$) & &  & ($10^{-3}\, \muHz^{-1}$) \\
\noalign{\smallskip}
\hline
\noalign{\smallskip}
11.126 & 5.65 & $0.0445$  &1.313  & 1.107  \\
11.131 & 5.38 & $0.0405$  & 1.358  & 1.114  \\
11.135 & 5.15 & $0.0373$ & 1.507   & 1.208 \\
11.139 & 4.97 & $0.0339$ & 1.631  & 1.244 \\
\noalign{\smallskip}
\hline
\end{tabular}
\begin{tablenotes}
\setlength\labelsep{0pt}
\scriptsize
\item \textbf{Notes.} {The coefficient $\beta$ represents the slope of the linear relation between mode frequencies and coupling factor, when equation~\eqref{eq:linq} is rearranged to the normal linear model of equation~\eqref{eq:linq1}.
}
\end{tablenotes}
\end{table}

\clearpage

\begin{figure*}
\resizebox{1.0\hsize}{!}{\includegraphics[angle=0]{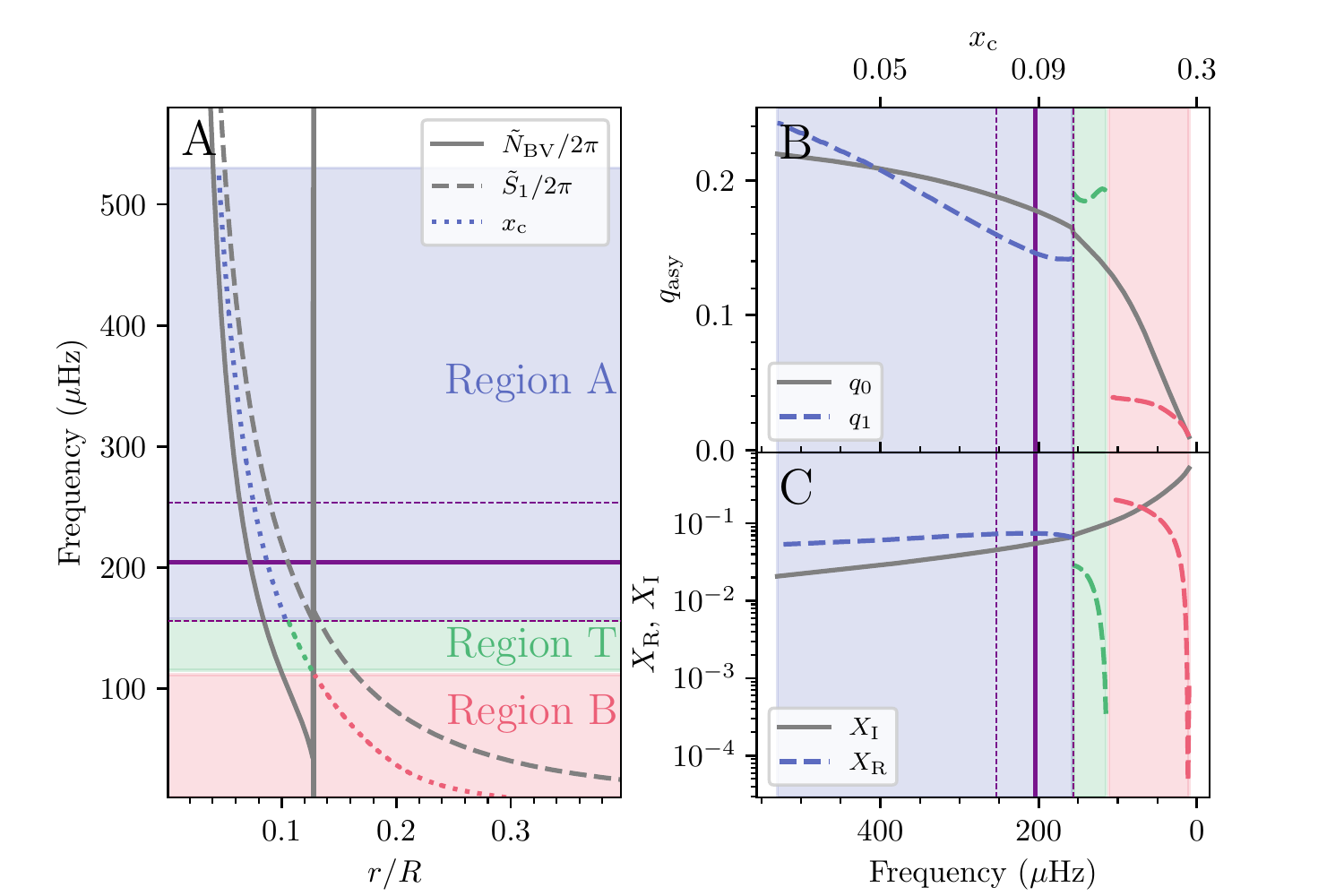}}
\caption{Illustrations for the younger model (Model 1, $\Dnu = 16.27\, \muHz$) in Figures~\ref{fg:propagation} and~\ref{fg:track}. The evanescent zone is divided into three regions, Region\,A (blue), Region\,T (green) and Region\,B (red), when taking the buoyancy spike into consideration. The purple dashed lines mark the observable frequency range with a width of $6~\Dnu$ that is centred by the frequency $\numax$ of maximum power (purple solid lines).
\textit{Panel A}: Propagation diagram showing the mid-layer structure near the base of the convection zone. The dotted line shows the central positions ($x_{\rm c}$) of the evanescent zone, as a continuous function of frequency (see main text in Section~\ref{sc:evo_asy}). \textit{Panel B}: $q_{\rm asy}$ as a function of frequency, computed with equation~\eqref{eq:weakq} for $q_0$ and  equation~\eqref{eq:qasy} for $q_1$. \textit{Panel C}: The gradient term $X_R$ and the integral term $X_I$ as a function of mode frequency, explaining the two significant jumps seen in $q_1$ at the borders of the Region\,T.}
\label{fg:q_vs_fre}
\end{figure*}

\begin{figure*}
\resizebox{1.0\hsize}{!}{\includegraphics[angle=0]{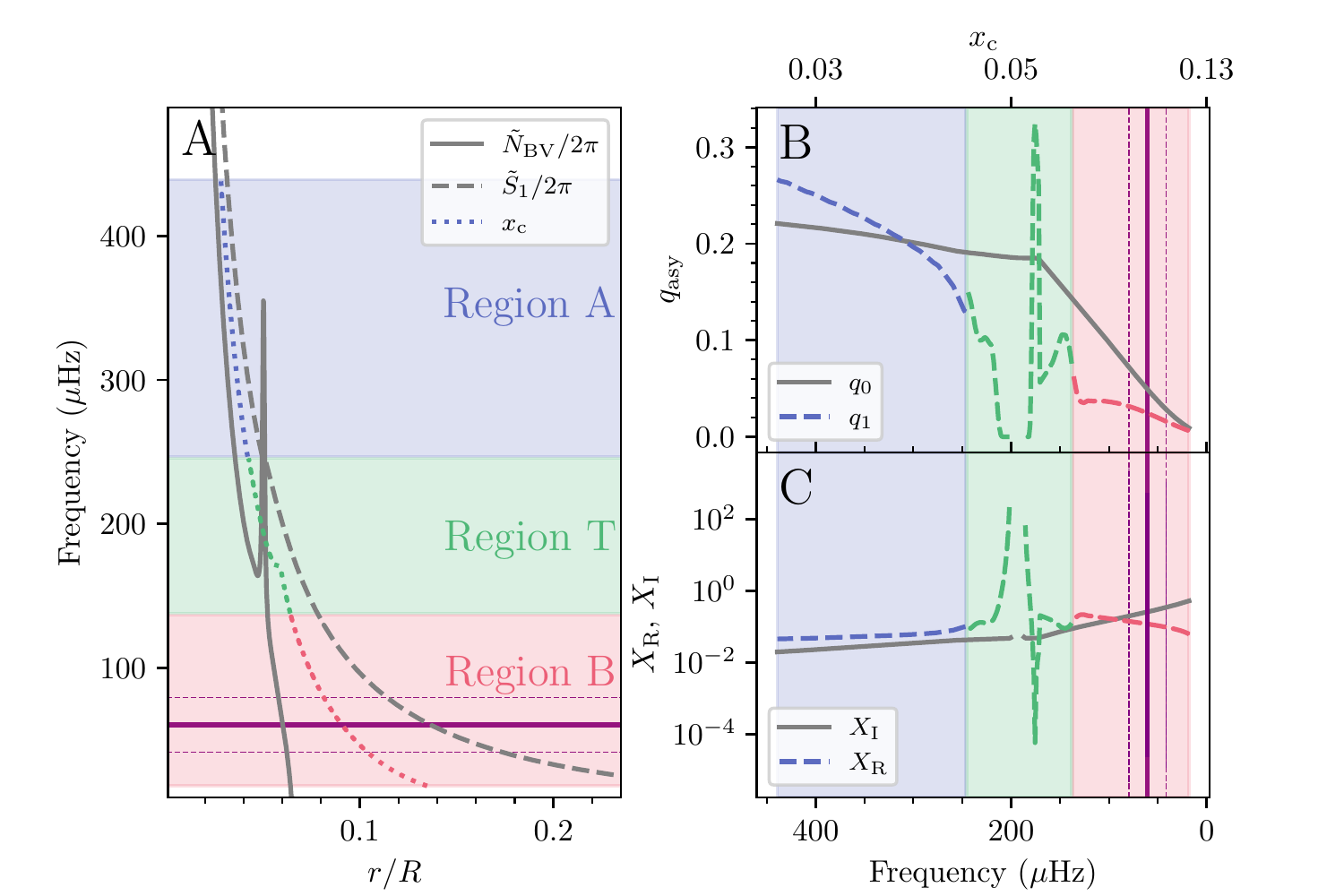}}
\caption{Illustrations for Model 2 ($\Dnu = 6.35\, \muHz$), using the same plot configurations as Figure~\ref{fg:q_vs_fre}.}
\label{fg:q_vs_fre1}
\end{figure*}

\begin{figure*}
\resizebox{1.0\hsize}{!}{\includegraphics[]{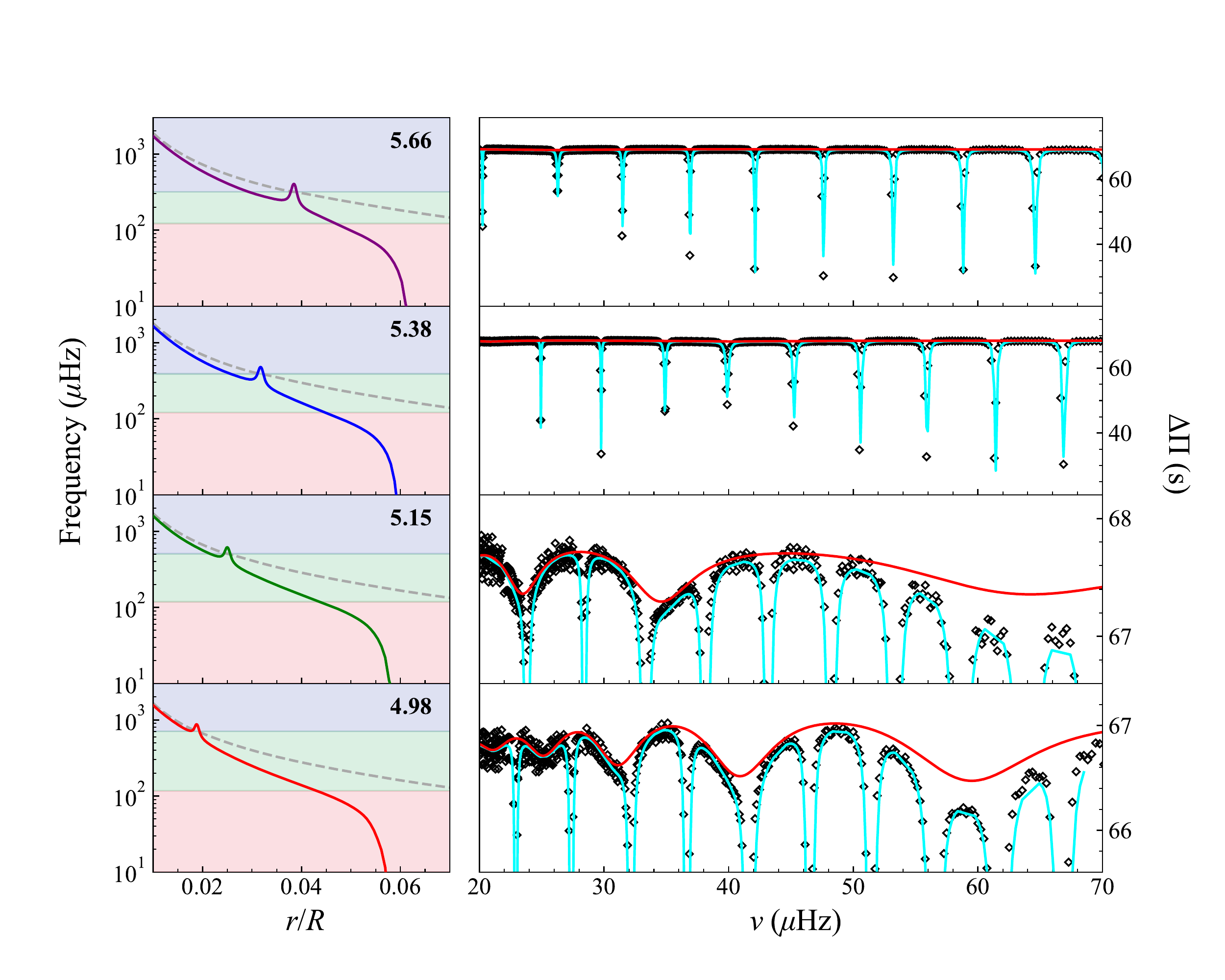}}
\caption{\textit{Left column}: $\tilde{N}_{\rm BV}$ (solid and coloured) and $\tilde{S}_{\rm 1}$ (dashed) as a function of fractional radius for the four RGB models (1.0 $\msun$, $Z=0.0173$) that have a core glitch. The values of $\Dnu$ (in unit of $\muHz$) are given at the upper right corner of each plot. The evolutionary stage of each model is shown in the inset plot of Figure~\ref{fg:track}, as the symbol colours matching the colours of $\tilde{N}_{\rm BV}$ here. We note that the model on the RGB bump positioned where the luminosity just starts to decrease has a $\Dnu=4.86 \, \muHz$. Regions\,A, T and B are shown in the same colour configuration as in Figures~\ref{fg:q_vs_fre} and \ref{fg:q_vs_fre1}. \textit{Right column:} comparisons between period spacings computed from {\scriptsize{ADIPLS}} (black diamonds) and those (cyan) derived from equation~\eqref{eq:glitch} using the fitted parameters listed in Table~\ref{tb:glitch}, for the same four models as in the left column. The red curves exhibit the impact of the buoyancy glitch on the period spacings, when the mode coupling is ignored. 
The bottom two panels show a zoom within a small range around $\Dpi_{\rm asy}$
to highlight the glitch impact.} 
\label{fg:glitch}
\end{figure*}

\begin{figure}
\resizebox{\columnwidth}{!}{\includegraphics[]{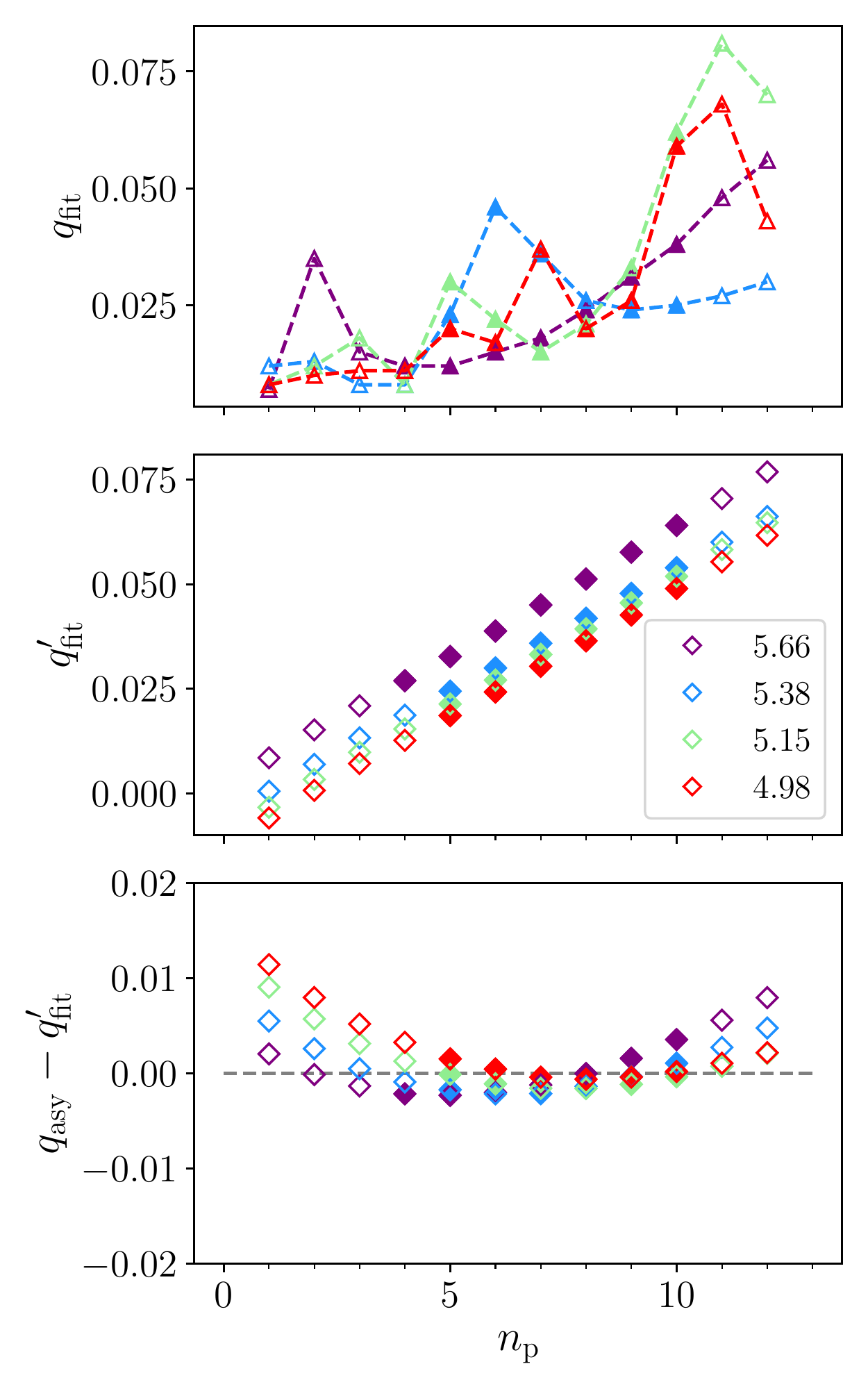}}
\caption{Coupling factors obtained from different methods as a function of $\np$ for the same four RGB models with a core glitch shown in Figure~\ref{fg:glitch}. The symbol colours imply the $\Dnu$ and thus the ages of different models. The values of $\Dnu$ implied by each colour are given in the legend. The top panel uses the same colour scheme as in the other two panels, though with different symbols. Detectable modes with frequencies within $\numax \pm 3 \Dnu$ are indicated by filled symbols. \textit{Top}: $q_{\rm fit}$ obtained by J20, based on equation~\eqref{eq:mixnu}, which does not account for the glitch impact. \textit{Middle}: $q^\prime_{\rm fit}$ derived from equation~\eqref{eq:glitch}, accounting explicitly for the glitch impact as well as coupling. \textit{Bottom}: the residuals ($q_{\rm asy}$ - $q^\prime_{\rm fit}$),  with $q_{\rm asy}$ derived from equation~\eqref{eq:weakq}. The dashed line is used to indicate agreement.}
\label{fg:qglitch}
\end{figure}

\clearpage

\appendix
\section{Asymptotic expression of $X$}
\label{app:x}
T16 gave the expressions of $X_I$ and $X_R$ in the asymptotic limiting case of a very thin evanescent zone.  Here we only present the important steps with associated quantities that are relevant to $X_I$ and $X_R$. For detailed derivations we refer to T16.

\cite{tak06} showed that a second-order system of differential equations can be used to describe the adiabatic dipolar oscillations as
\begin{equation}
r \frac{\mathrm d}{\mathrm d r}
\begin{pmatrix} \frac{\zeta_r}{r} \\ \frac{\zeta_h}{r} 
\end{pmatrix} = 
\begin{pmatrix} \mathcal{V} + J - 4 & 2J-\lambda \mathcal{V} \\ 
J - \frac{\mathcal A}{\lambda} & \mathcal{A} + 2J - 4 
\end{pmatrix}
\begin{pmatrix}
\frac{\zeta_r}{r} \\ \frac{\zeta_h}{r}
\end{pmatrix},
\label{eq:diffeq}
\end{equation}
in which the dependent variables, $\zeta_r$ and $\zeta_h$, are the reduced radial and horizontal displacements for dipolar oscillations, respectively. The other quantities on the right-hand side of equation~\eqref{eq:diffeq} are defined in terms of the pressure $p$, the first adiabatic index $\Gamma_1$,  the density $\rho$ and the gravitational acceleration $g$ by
\begin{equation}
\mathcal V = - \frac{1}{\Gamma_1 J} \frac{\mathrm{d} \ln p}{\mathrm{d} \ln r},
\end{equation}
\begin{equation}
\mathcal A = \frac{1}{J} \left( \frac{1}{\Gamma_1} \frac{\mathrm{d} \ln p}{ \mathrm{d} \ln r} - \frac{\mathrm{d} \ln \rho}{ \mathrm{d} \ln r} \right),
\end{equation}
and
\begin{equation}
\lambda = \frac{\omega^2 r }{g},
\end{equation}
where $\omega$ is the angular frequency and $J$ is provided by equation~\eqref{eq:j}. Then we can define two quantities
\begin{equation}
P = 2 J - \lambda \mathcal V
\label{eq:pr}
\end{equation}
and 
\begin{equation}
Q = J - \frac{\mathcal A}{\lambda}.
\label{eq:qr}
\end{equation}
Note that $\sqrt{-PQ}/r$ means the radial wavenumber in the p- and g- mode cavities in the asymptotic limit \citep[cf. sections 15 and 16 of][]{unno89}. 
The two characteristic frequencies are linked to $P$ and $Q$ through $\mathcal V$ and $\mathcal A$ as
\begin{equation}
\tilde N_{\rm BV}^2 = \frac{g}{r} \frac{\mathcal A}{J}
\end{equation}
and
\begin{equation}
\tilde S_1^2 = \frac{g}{r} \frac{2 J}{\mathcal V}
\end{equation}

All above mentioned quantities are functions of $r$. T16 rewrote the pulsation equations in terms of the new independent variable $s$ that is related to $r$ as
\begin{equation}
s = \ln x - \frac{1}{2} (\ln x_1 + \ln x_2) = \ln \frac{x}{\sqrt{x_1 x_2}},
\end{equation}
where $x$ is the fractional radius $r/R$, with $R$ being the radius of the star. 
Thus the turning points $x_1$ and $x_2$ are transformed to $s=s_0$ and $s=-s_0$, respectively, with $s_0$ given by
\begin{equation}
s_0 = \frac{1}{2} (\ln x_1 - \ln x_2) = \ln \sqrt{\frac{x_1}{x_2}}.
\end{equation}
Note that $s_0$ is negative when $x_1 < x_2$. Thus the centre of the EVZ is located at the point where $s = 0$ or $x = x_{\rm c}$, as shown in Figure~\ref{fg:q_vs_fre}.

$P$ and $Q$ are then rewritten as a function of $s$ by
\begin{equation} 
\frac{P}{F^2} = - S(s) (s-s_0)
\label{eq:ps}
\end{equation}
and
\begin{equation} 
Q{F^2} = T(s) (s+s_0),
\label{eq:qs}
\end{equation}
in which 
\begin{equation}
S > 0 ~~ \text{and} ~~ T > 0,
\end{equation}
and
\begin{equation}
F(x) = \exp{ \left[ \frac{1}{2} \int^x \frac{ V(x_{\rm a}) - A(x_{\rm a}) - J(x_{\rm a}) }{x_{\rm a}} \mathrm d  x_{\mathrm a} \right]}.
\end{equation}
The integral term $X_I$ in equations~\eqref{eq:weakq} and~\eqref{eq:qasy} is then given by
\begin{equation}
X_I = \frac{1}{\pi} \int_{-|s_0|}^{|s_0|} \kappa(s) \sqrt{s_0^2 - s^2} \, \mathrm d \, s,
\label{eq:xi}
\end{equation}
with
\begin{equation}
\kappa(s) = \sqrt{S T}  = \sqrt{\frac{-P Q} {s^2 - s_0^2}}.
\end{equation}
And the other gradient term $X_R$ is defined as
\begin{equation}
X_R = \frac{1}{2 \kappa (0)} \left(  \frac{\mathrm d \ln \mathfrak c}{\mathrm d s} \right)^2_{s=0},
\label{eq:xr}
\end{equation}
where 
\begin{equation}
\mathfrak c = \left(\frac{S}{T}\right)^{1/4}.
\end{equation}
\label{lastpage}
Thus, $X_R$ is sensitive to the variations of $\tilde N_{\rm BV}$ and $\tilde S_1$ in the vicinity of $x_{\rm c}$ where $s=0$.

\end{document}